\documentclass{aa}  

\usepackage{graphicx}
\usepackage{multirow}
\usepackage{txfonts}

\usepackage{placeins}
\usepackage{xspace}
\usepackage{amsmath}
\usepackage{mathtools}
\usepackage{xcolor}
\usepackage{float}

\newcommand{\starry}{\textsf{starry}\xspace}
\begin{document} 

    \title{Towards Doppler eclipse mapping of hot Jupiters}
    \subtitle{An observational perspective on WASP-33 b  with SPIRou}

   \author{V. Yariv \inst{1}
          \and 
          X. Bonfils \inst{1}
          \and
          N. Cowan \inst{2}
          \and
          T. Forveille \inst{1}
          \and
          F. Debras\inst{3}
          \and
          A. Carmona \inst{3}
          \and
          X. Delfosse \inst{1}
          \and
          A. Masson \inst{4}
          \and
          V. Parmentier \inst{5}
          \and
          R. Allart \inst{6}
          }

   \institute{Institut de Planétologie et Astrophysique de Grenoble, Grenoble, CNRS, IPAG, 38000 Grenoble, France \\
              \email{vincent.yariv@univ-grenoble-alpes.fr}
        \and
        Department of Earth \& Planetary Sciences, McGill University, 3450 rue University, Montréal, QC H3A  0E8, Canada 
        \and
        Institut de Recherche en Astrophysique et Planétologie, CNRS, UPS, Toulouse, France 
        \and
        Centro de Astrobiología (CAB), CSIC-INTA,  Camino Bajo del
Castillo s/n, 28692, Villanueva de la Cañada, Madrid, Spain 
        \and
        Université Côte d’Azur, Observatoire de la Côte d’Azur, CNRS, Laboratoire Lagrange, France
        \and
        Département de Physique, Institut Trottier de Recherche sur les Exoplanètes, Université de Montréal, Montréal, Québec, H3T 1J4, Canada
             }

   \date{Received XX, XX; accepted XX, XX}

 
  \abstract
   {In the last decade, ground-based high-resolution spectroscopy (HRS) has emerged as a powerful method to probe exoplanet atmospheres both in transit and thermal emission. With tens of instruments worldwide, HRS is now producing numerous observations on many targets, revealing these planet's thermal, compositional, and dynamical structure in three dimensions.  As HRS science continues to mature, novel strategies and interpretation tools will be key to extracting the maximal scientific output from these rich datasets.}
   {For this article we investigated the potential of exploiting eclipses in order to retrieve spatial constraints on exoplanet dayside atmospheres with HRS,  an approach that has been successfully applied at lower spectral resolutions with space-based facilities.}
   {To attempt this, we obtained an observation programme covering eight eclipses (ingress and egress) of the ultra-hot Jupiter WASP-33b using the (R$\rm\sim$70,000) SPIRou spectropolarimeter on the 3.6m CFHT. We analysed these data with the publicly available ATMOSPHERIX pipeline, which we combined with the \starry Python package to fit the eclipse mapping signal. Additionally, we performed injection-recovery tests on archival SPIRou data to evaluate the detection limits that could be reached with further observations.}
   {From eight ingresses and egresses of WASP-33b, we obtain a tentative detection of CO consistent with the literature values and archival SPIRou data, validating that eclipses may be stacked coherently to boost detection limits. In combination with longer phase-coverage dayside data, our eclipses marginally improve constraints on the planetary rotation velocity. Through injection recovery tests, we show that our results are scaling according to expectations for a synchronously rotating WASP-33b, implying that a further 15 eclipses with SPIRou ($\rm\sim$20 h) would be required to measure the planet's rotation using this method. While our study reveals the potential of HRS eclipses observations, it also highlights the importance of solving the remaining challenges in data processing for short time series and/or slowly accelerating planets ahead of the next generation of telescopes.}
   {}

   \keywords{Exoplanets -- High Resolution Spectroscopy -- Hot Jupiter
               }

   \maketitle
%

\section{Introduction} \label{sec:intro}

\begin{figure*}[ht!]
    \centering
    \includegraphics[width=0.8\textwidth]{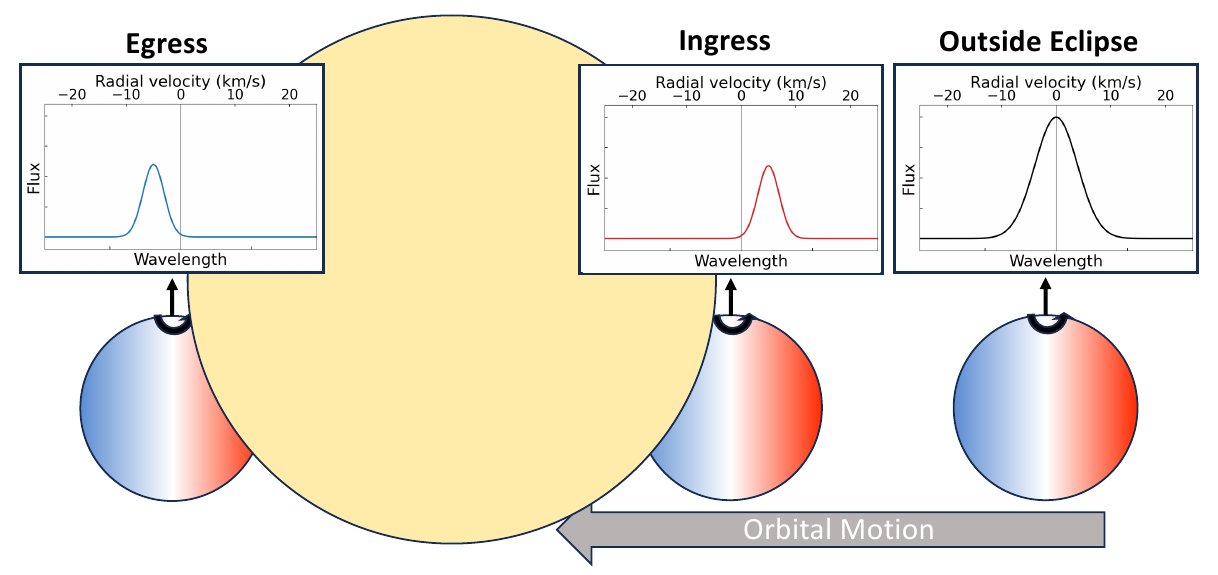}
    \caption{Schematic representation of the changing shape of an emission line during secondary eclipse. During ingress--egress, the rotation produces an offset in the line centroid, rather than just symmetrical line broadening.}
    \label{fig:Line_Shape}
\end{figure*}

Boasting large radii and high equilibrium temperatures, hot Jupiters (HJs) have proven to be extremely fruitful exoplanets for in-depth atmospheric characterisation. Space-based spectroscopic observations have enabled detailed measurements of HJ atmospheric compositions and temperature distributions, revealing strong day--night contrasts shaped by extreme irradiation and synchronous rotation (e.g. \citealt{mansfield_hstwfc3_2018,mikal-evans_confirmation_2020}). Transit spectroscopy, which provides stronger spectral features, is particularly powerful for probing compositions in the upper atmosphere at the day--night terminator, while emission spectroscopy can probe compositions and temperatures deeper into the atmosphere and across the entire planet. Eclipses, where the planet is progressively hidden behind its star, provide a unique opportunity to spatially resolve its dayside \citep{rauscher_toward_2007,majeau_two-dimensional_2012,de_wit_towards_2012}. Recently, the James Webb Space Telescope (JWST) has enabled spectroscopic phase curves at unprecedented precisions, leading to the first eclipse maps that reveal 3D thermal structure on ultra-hot Jupiters (UHJs) \citep{coulombe_broadband_2023,challener_horizontal_2025}.

In parallel, exoplanet characterisation from ground-based telescopes has emerged as a competitive and complementary alternative to space-based observations, through the use of high-resolution spectroscopy (HRS; see reviews by \citealt{birkby_exoplanet_2018, snellen_exoplanet_2025}). By resolving individual atomic or molecular lines, HRS can detect the Doppler shift from the planet's orbit at a precision of a few kilometres per second, and thus separate its signal from the relatively static stellar and telluric lines. Typically, constraints are retrieved via cross--correlation with model planet spectra, leveraging thousands of lines to overcome the stellar noise and to robustly detect the signature of a given species \citep{zucker_cross-correlation_2003,brogi_retrieving_2019}. Much like its space-based counterpart, HRS study of exoplanet atmospheres has initially focused on HJs and UHJs, which provide the most favourable signal. As detections have become commonplace, the next frontier for HRS lies in developing robust frameworks to convert the compositional and kinematic information encoded in these datasets into a three-dimensional description of the structure and dynamics of exoplanet atmospheres to compare with detailed physical models. 

In transmission, several observations have already revealed velocity offsets between species indicative of vertically stratified wind and/or abundance patterns on HJs \citep{flowers_high-resolution_2019,ehrenreich_nightside_2020,cauley_time-resolved_2021,hood_atmospherix_2024,seidel_vertical_2025}, consistent with predictions from global climate models (GCMs) \citep{showman_atmospheric_2002,showman_equatorial_2011}. In emission, where the signal-to-noise ratio (S/N) tends to be lower, such measurements remain more challenging to interpret. Multiple observations have found velocity offsets between various atoms or molecules, indicative of inhomogeneous contributions of these species across observed UHJ daysides (e.g. \citealt{cont_silicon_2022,bazinet_quantifying_2025}). Line broadening has also been used to constrain the bulk rotation of these planets (e.g. \citealt{brogi_rotation_2016,boucher_co_2023,lesjak_retrieving_2025}), but both these effects remain difficult to interpret robustly and consistently as they are degenerate with the spatial distribution of the observed species. Much as they have been used from space, eclipses could thus provide a promising way to break this degeneracy and measure the rotation and winds on HJ daysides. 

For this paper we explored the possibility of exploiting HRS eclipse observations of (U)HJs to measure their rotation and atmospheric dynamics, and present initial results from the first pilot study attempting this on the UHJ WASP-33b. In section \ref{sec:Models}  we introduce the notion of Doppler eclipse mapping through the reciprocal Rossiter-McLaughlin effect, before presenting our methods to model the associated velocity anomalies allowing us to identify promising targets. In section \ref{sec:sig_obs}  we discuss the SPIRou observations we obtained for this study on our most promising target WASP-33 b. Section \ref{sec:Analysis} describes our methods for data processing, model fits, and injection-recovery tests; the results of which are presented in section \ref{sec:Results}. Finally, section \ref{sec:Discussion} provides some discussion points on these results and future perspectives, with concluding remarks in section \ref{sec:concl}.\\

\section{Eclipse mapping at high resolution}\label{sec:Models}
\subsection{RM effect at secondary eclipse}\label{sec:RMse}

When a rotating body is occulted, the specific radial velocity (RV) components originating from the hidden portion of its surface are removed from its spectrum, resulting in an apparent line-shifting effect. First observed by \citet{rossiter_detection_1924} and \cite{mclaughlin_results_1924} in eclipsing binaries, this phenomenon is commonly known as the Rossiter-McLaughlin (RM) effect. The RM effect is extensively studied on stars during exoplanetary transits, as the shape and amplitude of the RV anomaly can be used to recover stellar rotation velocities, axial tilts and spin-orbit alignments with the transiting planets \citep{winn_transits_2014,triaud_rossiter-mclaughlin_2018}. 

When the planetary companion is occulted by its host star during secondary eclipse, a similar line-shifting effect is expected on the planet's velocity (see Fig. \ref{fig:Line_Shape}). 
Limited to ingress and egress -- beyond which the planet is fully hidden -- this reciprocal RM effect could provide a unique way to constrain rotation rates and axial tilts of exoplanets. Planets being much fainter than their host stars, this effect is much more challenging to detect than on stars, and has thus garnered little attention in the literature. The single study known to the authors at the time of writing is presented in \cite{nikolov_radial_2015}, who provide a formal description of the effect, and show how it can be used to recover planetary rotation rates and axial tilts. They finally estimate observational perspectives by applying scaling laws to the water detection in HD189733b from \cite{birkby_detection_2013}, concluding that a feasible programme to detect this effect would require several eclipses with a 40m class telescope.

Since 2015, exoplanet science and HRS have greatly matured: the TESS mission has uncovered many new potential targets, while high S/N detections in HRS have been reported for a much wider range of planets and species. These advancements, along with the approaching prospect of extremely large telescopes (ELT), provide strong motivation to revisit the potential of eclipses with HRS. In the sections that follow, we describe our approach to modelling the planetary RV anomaly during eclipse (hereby referred to as eclipse mapping signal), and apply it to the known population of UHJs to identify the most promising targets based on photon noise statistics.

\begin{figure*}[ht!]
    \centering
    \includegraphics[width=0.96\textwidth]{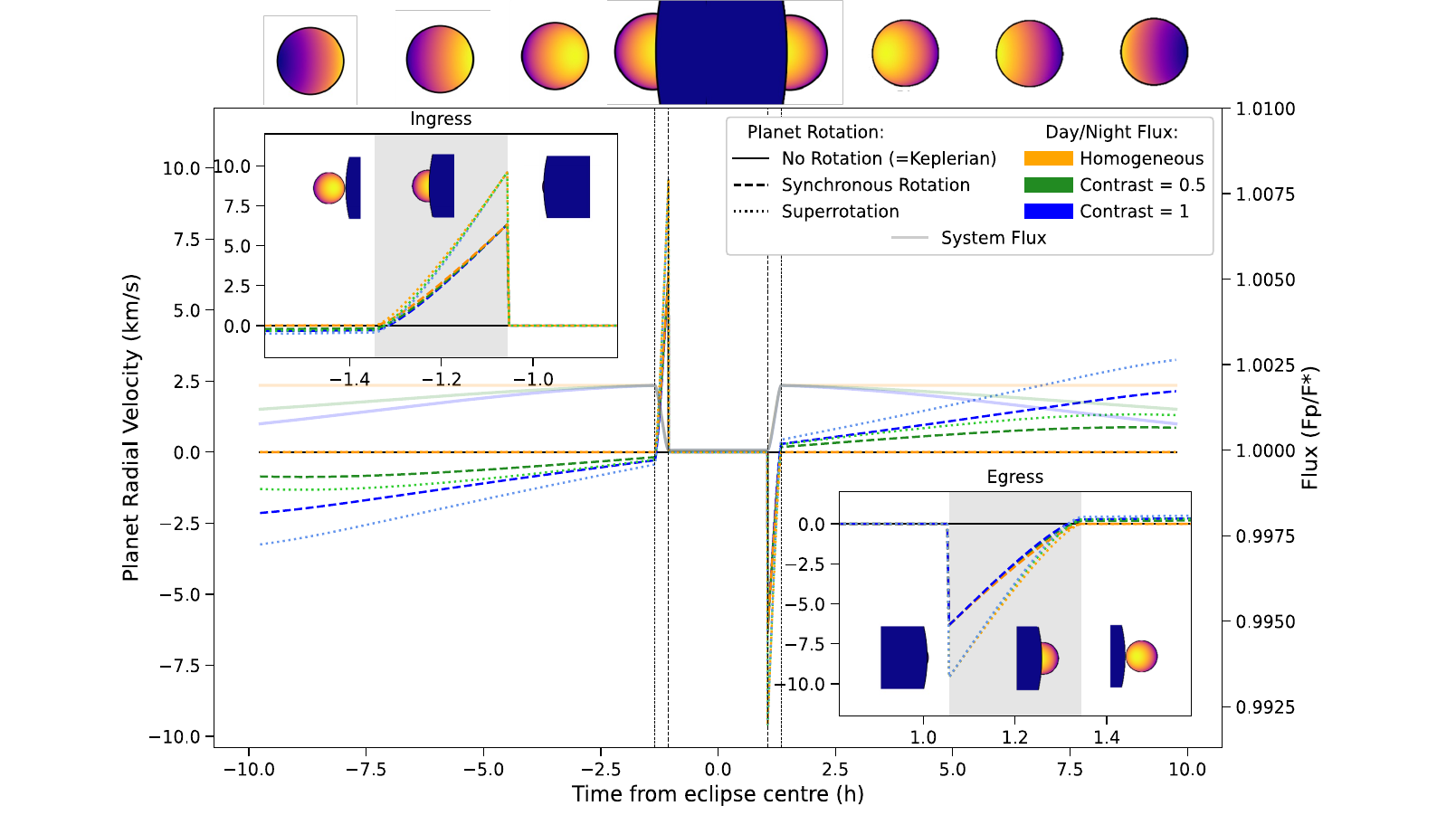}
    \caption{Radial velocity offset of WASP-33 b from a Keplerian orbit due to the rotation effect for different planet intensity maps and rotation velocities. The insets show zoomed-in images of the occultation ingress and egress phases. The total flux is plotted on the secondary axis for the three day--night contrast values considered. As can be seen, the RV offsets are both more pronounced and less sensitive to the day--night contrast vs during outside occultation, validating that eclipses would be a promising and complementary approach to more classic dayside observations for constraining rotation velocities.}
    \label{fig:starry_outputs}
\end{figure*}

\subsection{Modelling planetary velocity anomalies}\label{sec:starry}

To compute velocity offsets caused by the planetary rotation during eclipse, we used the Python package \starry \citep{luger_starry_2019}. \starry enables fast and accurate computations of phase and occultation light curves for objects whose surface intensity can be expressed as a sum of spherical harmonics. Specifically, we used \starry's ability to compute intensity-weighted radial velocities for solidly rotating spherical bodies \citep{luger_mapping_2021}. For each system we studied, we created a \starry model with the corresponding stellar, planetary, and orbital parameters. The star being the occulting body, its surface structure is irrelevant and we gave it a uniform intensity of $F_*/F_p$, while the planetary intensity was set to 1. We parametrised the planetary surface intensity with a first-degree spherical harmonic $Y_{l=1,m=0}$, such that the intensity is monotonically from the darker nightside hemisphere to the brighter dayside.\footnote{\starry defines the $Y_{l=1,m=0}$ harmonic longitudinally rather than latitudinally.} This intensity map, which is used as a weight on the radial velocities from each point on the surface, needs to remain positive or null, which we achieved by setting $0 \leq Y_{1,0} \leq 1/\sqrt{3} $. One can at this point introduce a phase offset to our planet and/or consider an obliquity of its rotation axis. For the purpose of target selection, we chose to limit our exploration to simple cases with no offset and an obliquity of 90 degrees, i.e. the most optimistic scenario for detection. While our SPIRou data (see Sect. \ref{sec:sig_obs}) are not constraining enough to consider more complex cases in our analysis, we discuss the impact of these other parameters in section \ref{sec:Discussion}.

The \starry modelling provides two outputs, which we combined with a template planetary spectrum (see section \ref{sec:Templates}) to generate a time series including rotational effects both in and out of eclipse. The first output is a flux curve for the integrated visible day or night portion of the planetary surface, which we used to scale the template spectrum's intensity at each orbital phase. The second output is the planet’s intensity-weighted radial-velocity anomaly, $\Delta RV$, i.e. the rotation-induced deviation from a simple Keplerian velocity, which we applied as a phase-dependent spectral shift and line broadening as described further in section \ref{sec:sig_inj}. Figure \ref{fig:starry_outputs} shows examples of flux curves and velocity anomalies for WASP-33 b (see sections \ref{sec:target},\ref{sec:wasp-33}) under different rotation rates and day–night contrast scenarios. Outside eclipse, velocity anomalies arise from the changing visibility of the day and night hemispheres, and thus depend strongly on the brightness contrast. In contrast, RV anomalies in ingress and egress are larger, evolve on a much shorter timescale, and are comparatively insensitive to the day–night contrast. Consequently, beyond constraining rotation rates and obliquities, the eclipse mapping signal could help to break degeneracies between spectral maps and the out-of-eclipse velocity offsets.

\subsection{Target selection}\label{sec:target}

Armed with the modelling tools described above, we can evaluate theoretical detection limits for the eclipse mapping signal across a broad sample of targets and instruments. We focused on UHJs,\footnote{Selected from the NASA Exoplanet Archive by requiring planetary radii $\geq 1,R_\mathrm{J}$ and equilibrium temperatures above 2000 K.} as their high dayside temperatures and thermal inversions are expected to yield stronger high-resolution emission features. We also decided to focus our study specifically on the eclipse mapping signal obtainable from CO lines in the K band, as these have led to some of the strongest detections in HRS, thanks to the more favourable planet to star flux contrast in the infrared. Moreover, CO has been shown to be a very good probe of atmospheric dynamics, thanks to its strong spectral lines and thermal stability across UHJ daysides \citep{van_sluijs_carbon_2023,wardenier_pre-transit_2025}.

Based on the system parameters obtained from the NASA Exoplanet archive on our sample, we ran \starry simulations of eclipse ingress--egress for each system assuming synchronous rotation, and generate a template spectrum as described in section \ref{sec:Templates}, where we rescaled the temperatures in the T-P profile with each planet's equilibrium temperature. The resulting spectra were shifted and broadened phase-dependently using the velocity offsets computed from \starry as described in section \ref{sec:sig_inj}. The spectral intensity was then converted to photon counts based on a stellar blackbody, flux ratio $F_p/F_*$, and distance in parsecs to the system. Finally, the spectra were broadened with a Gaussian kernel to account for instrumental broadening, summed over an entire ingress--egress and scaled for a given telescope diameter and overall instrument efficiency. Using these rescaled model spectra, we computed the photon-noise limit on the precision of the planet’s radial velocity (see Appendix \ref{app:photon}) achievable by a given instrument. Comparing these limits with the expected velocity anomalies modelled with \starry, we can thus assess the detectability of the eclipse mapping signal of each planet--instrument pairing. 

As opposed to a simple emission spectroscopy metric, which depends mainly on the system magnitude and planet-to-star flux ratio, the eclipse mapping detectability is a more complex relationship between the various planetary and instrumental parameters. Broadly, in addition to a good emission metric, the important parameters for a high eclipse mapping detectability are a fast rotation velocity, which increases the amplitude of the eclipse mapping signal, and a long ingress--egress duration, which increases the S/N that can be reached in a single event. Since we consider only tidally locked rotation, a faster rotation implies a shorter orbital period,\footnote{and/or larger radius} but this results in shorter ingress--egress duration. Planets with a high impact parameter (e.g. TOI-1518b), can thus be particularly promising, able to offer both a fast rotation and longer ingress-egress durations. In terms of instrumental parameters, higher resolutions ($\gtrsim$100K) are particularly beneficial for slower rotators $\rm<$5km/s, but can be de-emphasised in favour of a larger throughput at more moderate resolutions ($\sim$50-80K) for fast-rotating planets (5-10 km/s).

\begin{figure}[h!]
    \centering
    \includegraphics[width=\linewidth]{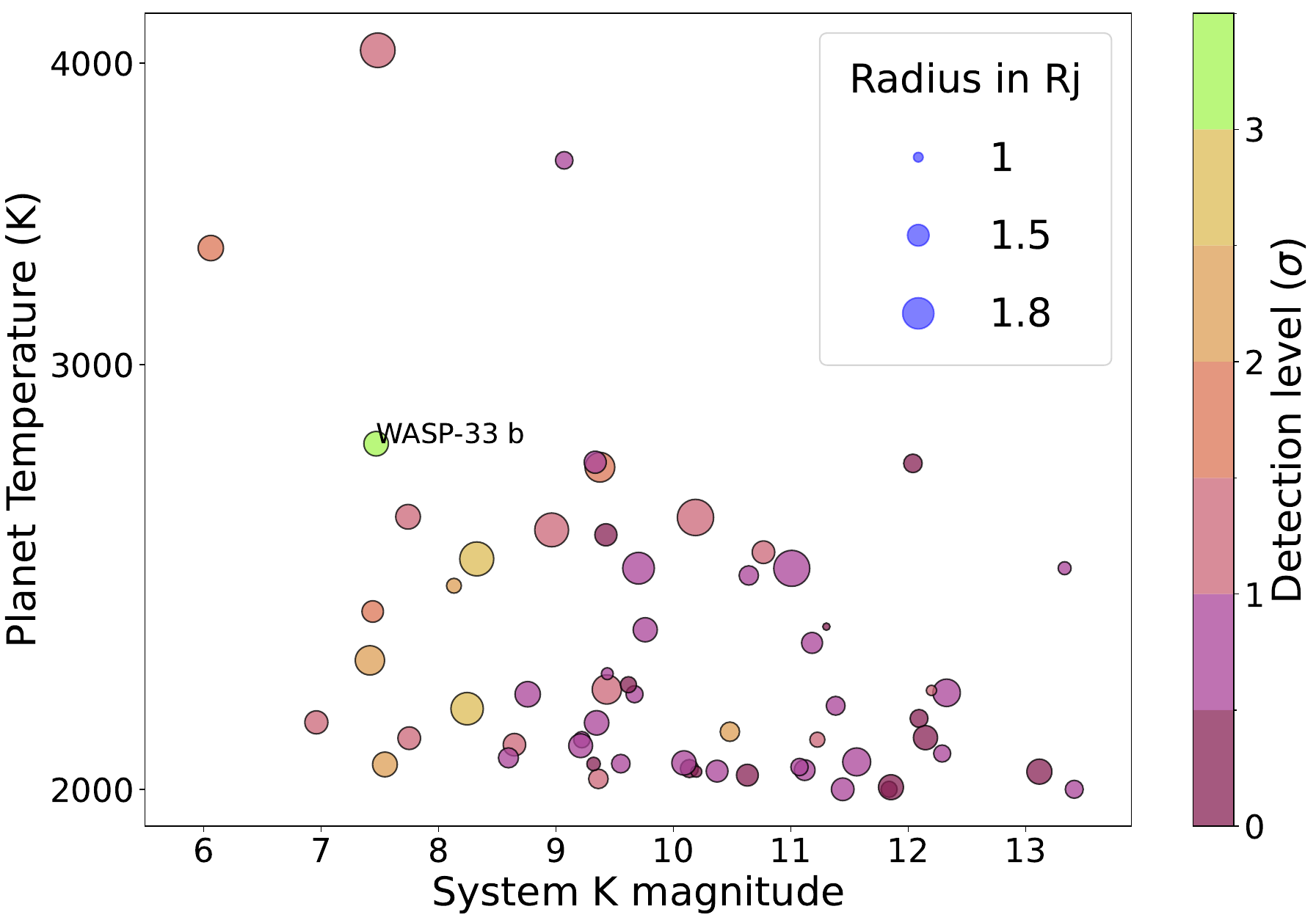}
    \caption{Photon-limited detection S/N of the radial velocity offset of CO lines induced by tidally locked rotation rates for a population of UHJs for ten occultations using instrument parameters for SPIRou. WASP-33 b is the only target for which the eclipse mapping signal could be detectable above the $3\sigma$ level.} \label{fig:SNR_RV}
\end{figure}

Overall, while we broadly agree with \citet{nikolov_radial_2015} that Doppler eclipse mapping will require the ELTs for most systems, our results are slightly more optimistic with regards to what can be achieved on current facilities. Figure \ref{fig:SNR_RV} shows the eclipse mapping detectabilities we obtain for the known population of UHJs with SPIRou over 10 occultation ingresses and egresses. WASP-33 b emerges as the best target, thanks to its favourable planet–star contrast (e.g. relative to KELT-9 b, WASP-18 b) and rapid rotation\footnote{V$\rm_{eq}$=6.7 km s$^{-1}$ assuming tidal locking} (e.g. relative to WASP-189 b, MASCARA-1 b). In this configuration, it is the only system for which the eclipse mapping signal is expected to reach the 3$\sigma$ level within $\sim$10 eclipses ($\sim$12 hours), making it uniquely promising for a pilot study with current facilities. Motivated by this, we obtained a programme to observe 8 eclipses with SPIRou, described in the following section.

\section{Observations} \label{sec:sig_obs}

We obtained eclipse (ingress and egress) observations of WASP-33 b with SPIRou in 2023 and 2024, which we complemented with archival dayside (pre- and post-eclipse) data from 2019. In order to evaluate detection limits for the eclipse mapping signal, we additionally injected a synthetic planet spectrum into these data, simulating further ingress--egress observations (see section \ref{sec:sig_inj}). In this section, we first summarise the properties of WASP-33~b and then describe the observations used in this work.

\subsection{WASP-33 b}\label{sec:wasp-33}

 WASP-33 b \citep{cameron_line-profile_2010} is a 1.6 $R_J$ planet orbiting its A-type host star with a period of 1.2 days. With an effective temperature of 2800 K, it is firmly within the population of UHJs, meaning it is expected to present a strong dayside hotspot and thermal inversion. This has been confirmed by both space-based and HRS observations, revealing large phase variations \citep{zhang_phase_2018} and line profiles in emission on the dayside (e.g. \citealt{nugroho_detection_2020,yan_detection_2022,cont_silicon_2022}). At high resolution, WASP-33 b has proven to be a particularly fruitful target, with multiple detections of atomic species including ionised calcium \citep{yan_ionized_2019,yang_evidence_2024} and hydrogen \citep{yan_detection_2021,cauley_time-resolved_2021,borsa_gaps_2021,yang_evidence_2024} in transmission, and iron \citep{nugroho_detection_2020,cont_detection_2021,herman_dayside_2022}, silicon \citep{cont_silicon_2022}, titanium, and vanadium \citep{cont_atmospheric_2022} in emission. More recently, detections of several molecular species have also been reported in emission, including CO \citep{yan_detection_2022,van_sluijs_carbon_2023,finnerty_keckkpic_2023,mraz_out_2024}, OH, and H2O \citep{nugroho_first_2021,finnerty_keckkpic_2023,choi_early_2025}. 

 In several cases, these detections have been interpreted in relation to the 3D properties of the planet. For instance, \cite{cont_atmospheric_2022} show that their lower-than-expected orbital velocity can be explained by the planetary rotation, which effectively blue- and redshifts the signal from the planetary dayside pre- and post-eclipse respectively (see Figure \ref{fig:starry_outputs}). In \cite{herman_dayside_2022}, the authors explicitly fit a brightness variation parameter, finding a positive phase offset of 22 degrees in their Fe detection, which they interpret as indicative of a westward hotspot offset. Through comparison with outputs from GCM simulations \cite{van_sluijs_carbon_2023} caution however that HRS phase offsets may not be directly mappable to temperature maxima, showing instead that their own weak preference for a positive phase offset in CO emission is best explained by an eastward hotspot offset. This interpretation would then also reconcile these HRS in results with Spitzer space telescope observations of WASP-33 b, which revealed an eastward-shifted hotspot \citep{zhang_phase_2018}.

 It is important to note finally that WASP-33 is a Delta Scuti variable star, which has hindered several attempts to detect molecular species in transit (e.g. \citealt{herman_search_2020,kesseli_search_2020}). The effects of stellar pulsations have also been noted in dayside observations of WASP-33 b, in particular for observations in the optical targeting atomic species, such as iron, which are likely to be present in the star. Some investigations (eg.\citealt{herman_dayside_2022}), opt to exclude any phases where planetary orbital and stellar rotation velocities overlap, in order not to bias their detections. In contrast, near-infrared observations targeting molecular species which are absent from the host star seem to be relatively weakly impacted by pulsations (e.g. \citealt{van_sluijs_carbon_2023}), which provides further grounds for targeting CO as a robust atmospheric tracer in this case, as we are exclusively observing close to secondary eclipse where these velocities overlap.

\begin{table}[h!]\label{tab:params}
\centering
\caption{WASP-33 system parameters}
\begin{tabular}{ccc}
\hline \hline
Parameter                          & Value             \\ \hline
\multicolumn{2}{l}{\textbf{Star Parameters}}           \\ \hline
Star Mass (M$_*$) *                & 1.561 $M_\odot$ $^a$      \\
Star Radius (R$_*$) *              & 1.509 $R_\odot$ $^a$      \\
Effective Temperature (T$_{eff}$)  & 7430 $\pm$ 100 K $^b$     \\
Rotation Velocity (v sin i$_*$ )   & 90 $\pm$ 10 km $s^{-1}$ $^b$  \\
Systemic Velocity (V$_{sys}$)      & $-$3 km $s^{-1}$  $^e$        \\
RV Semi-Amplitude (K$_*$)          & 304 $\pm$ 20 m $s^{-1}$ $^c$  \\ \hline
\multicolumn{2}{l}{\textbf{Planet Parameters}}         \\ \hline
Planet Mass (M$_p$) *              & 2.16 $\pm$ 0.2 $M_J$ $^c$ \\
Planet Radius (R$_p$) *            & 1.679 $R_J$ $^a$          \\
Orbital Period (P) *               & 1.21987089 days $^d$      \\
Epoch of Transit (T0) *            & 2456217.48712 BJD $^d$  \\
Semi-Major Axis (a)                & 3.69 $\pm$ 0.05 $R_*$ $^c$\\
Inclination (i) *                  & 86.2 $\pm$ 0.2 $^\circ$ $^a$\\
Eccentricity (e) *                 & 0                         \\
Orbital Semi-Amplitude (K$_p$) \textdagger    & 231 km $s^{-1}$           \\
Synchronous Rotation ($V_{eq}$) * \textdagger  & 6.7  km ${\rm s^{-1}}$ \\ \hline

\multicolumn{2}{l}{Velocities in barycentric rest frame unless otherwise stated}\\
\multicolumn{2}{l}{* Parameters used in \starry \textdagger Derived values} \\ 
\multicolumn{2}{l}{\textbf{References.} $^a$ \cite{kovacs_comprehensive_2013}; $^b$ \cite{cameron_line-profile_2010};} \\ 
\multicolumn{2}{l}{$^c$ \cite{lehmann_mass_2015}; $^d$ \cite{smith_cheops_2025}}; \\ 
\multicolumn{2}{l}{$^e$ \cite{nugroho_first_2021}}

\end{tabular}
\end{table}

\subsection{Observations}

SPIRou \citep{donati_spirou_2020} is a cryogenic high-resolution echelle spectrograph and polarimeter in the near-infrared, mounted on the 3.6m Canada-France-Hawaii Telescope (CFHT). It boasts a broad, continuous wavelength coverage ranging from 0.9 to 2.5$\mu m$, split over 49 overlapping diffraction orders (numbered 30 to 79 on the detector). In combination with its resolving power $\lambda/\Delta\lambda \approx 70 000$ ($\sim$4.29 km/s), these characteristics make SPIRou an ideal instrument for HRS of exoplanet atmospheres.

\begin{figure}[h!]
    \centering
    \includegraphics[width=\linewidth]{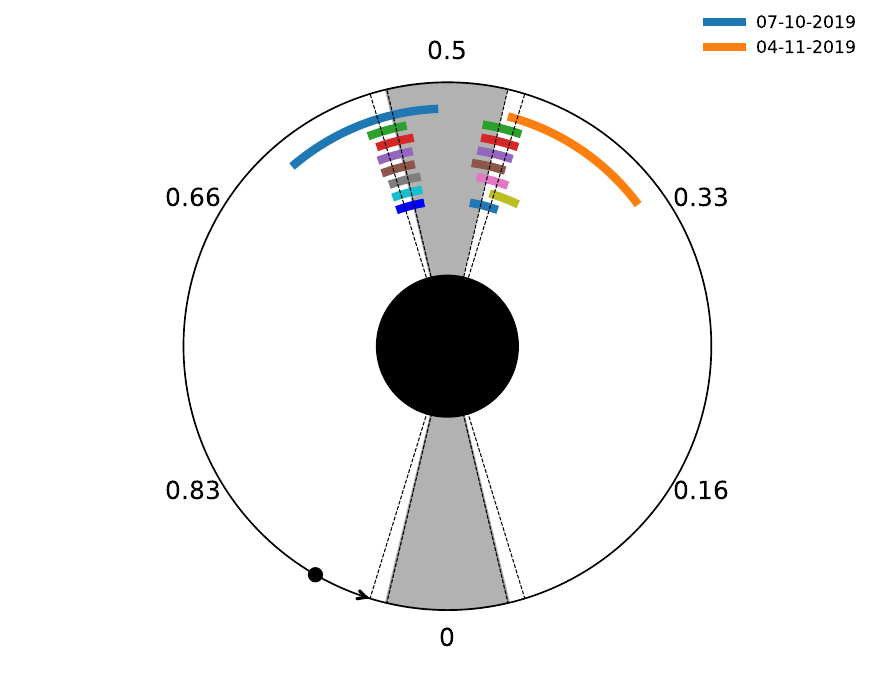}
    \caption{Phase coverage of the observations of WASP-33 b used in this work. The shaded regions represent transit--eclipse phases where the full planet disk is in front of or behind the star, while the dashed lines indicate the starts and ends of ingress and egress. The 2019 programme (PI: Darveau-Bernier), covering longer dayside phases, was taken from the CADC archive. Ingress and egress observations were obtained through the ATMOSPHERIX Large Programme (PI: F. Debras) in winter 2023 and a PI programme we obtained in winter 2024 (PI: N. Cowan). The full detail of all observations are given in Table \ref{tab:obs}.}
    \label{fig:phase_cov}
\end{figure}

Archival data from previous observations of WASP-33b's dayside taken between October and November 2019 (PI Darveau-Bernier, programme 19BC026) are publicly available in the SPIRou archive.\footnote{https://www.cadc-ccda.hia-iha.nrc-cnrc.gc.ca/} These cover several hours of WASP-33b's dayside to either side of secondary eclipse (see Fig. \ref{fig:phase_cov}), and are presented more fully in \cite{darveau-bernier_caracterisation_2023} and Darveau-Bernier et al. (in prep). In addition, we obtained several observations during occultation ingress and egress only: data for three ingress--egress pairs were obtained through the ATMOSPHERIX Large Programme (PI: F. Debras) in December 2023 (programme 23B), while another seven sets were obtained through a general observer programme (PI: N. Cowan) in August to December 2024 (programme 24B). The observations from programme 23B  were composed of eight exposures of 245s each in non-polarimetric mode, covering a full ingress--egress in four exposures including overheads. Due to a scheduling error, observation timings were slightly off during this programme. As such, only partial ingresses were obtained, while the egress observations were missed. For programme 24B, each sequence was extended to ten exposures in order to increase the in-eclipse baseline and ease up scheduling. Figure \ref{fig:phase_cov} shows the phase coverage of the observations from the archival programme and programme 24B, while a full summary of all the observations used can be found in Table \ref{tab:obs}. 

The observations were reduced using version 0.7.291 of the APERO pipeline \citep{cook_apero_2022}, the official data reduction software for SPIRou. The APERO pipeline extracts the spectrum from the H4RG detector using the optimal method described in \cite{horne_optimal_1986,artigau_h4rg_2018}. The wavelength solution is obtained by combining exposures from a UNe hollow-cathode lamp and a thermally stabilised Fabry-Perot etalon \citep{bauer_calibrating_2015,hobson_spirou_2021}. APERO also corrects for the telluric absorption following the method described in \citet{cook_apero_2022} and Artigau et al. (in prep). This method combines TAPAS atmospheric models and a library of observations of hot stars in different atmospheric conditions built over the lifetime of the instrument to accurately model the telluric absorption. APERO thus provides a time series of wavelength-calibrated telluric-corrected spectra which we used as input for further data analysis.

\section{Data analysis}\label{sec:Analysis}

\subsection{Data processing}\label{sec:data_proc}

Additional data processing is required to clean the spectrum, removing the stellar contribution and any remaining correlated residuals from telluric or instrumental conditions. We performed this processing with the publicly available ATMOSPHERIX pipeline \citep{klein_atmospherix_2024,debras_atmospherix_2024}, developed by the consortium to extract planetary spectra from high-resolution observations with SPIRou. To cover CO lines in the K band, we selected orders 32 and 33 ($\rm\lambda$=2.29-2.43 $\rm\mu$m)\footnote{Orders 31 and 34, also in the K band, were discarded as found to mainly contribute noise to the final CCF, due to few CO lines and high thermal and atmospheric noise, respectively} from the APERO-reduced spectra, on which we masked out any telluric lines with a transmission lower than 30\% in the core up to a transmission level of 95\% in the wings, following the method of \cite{boucher_characterizing_2021}.  We then applied the following steps on each order following the ATMOSPHERIX framework: \\

1. Shift the spectra to the stellar rest frame and create a master star spectrum for each observation night by taking the median of all exposures in the time series. This reference star spectrum is then shifted back to the geocentric frame, linearly matched in flux and divided out from each exposure. The same procedure is applied a second time in the Earth's rest frame to correct for tellurics.

2. Normalise each of the residual spectra with a rolling window mean filter, after which outliers above $5 \sigma$ are removed. This is repeated until no more outliers are left in the data. 

3. Correct for pixels with high temporal variation through an iterative parabolic fit on each pixel in the time domain, removing any outliers further than $5 \sigma$ from the fit.

4. Correct any remaining time-correlated noise per wavelength bin through a principal component analysis (PCA). The number of components to remove is scaled automatically for each order by applying a first PCA to a set of white noise maps scaled to the data's empirically estimated photon noise, saving the one with the maximal eigenvalue S$_{max}$. PCA is then applied to the observed spectra, removing 1 or any number of principal components with a significantly larger eigenvalue than S$_{max}$, whichever amount is larger (see \citealt{klein_atmospherix_2024}). 

For emission datasets, step (1) relies on the assumption that the planet signature is averaged out over many pixels due to its large Doppler shift over the course of the observation. In the case of the ingress--egress datasets, this assumption is no longer valid as the planet's Doppler shift moves it by only a few pixels over the short time series, and we thus skipped this step in the programme 23B and 24B  datasets.\footnote{We also tested creating master-spectra from the in-eclipse spectra only (i.e. when the planet is fully hidden by the star), but find this to perform worse than no master at all, likely because we have too few in-eclipse exposures per time series to create a good enough master-spectrum} To compensate for this, we decreased the filter size in step (2) for these observations from 150 to 50 pixels (i.e. $\sim$300 km/s down to $\sim$100 km/s), which we find performs slightly better in injection-recovery tests when not removing a master-spectrum. In step (4), the PCA auto-tuning resulted in 1 to 3 PCs being taken out per order in the K band for the observations in this study, with the shorter eclipse observations requiring fewer PCs than the dayside observations from programme 19B. Much like the master-spectrum step, the PCA cleaning is reliant on the important Doppler shift of the planetary spectrum, which can be kept relatively intact while the residual, comparatively fixed variations from the stellar, telluric and/or instrumental effects are effectively removed. In the case of a slow moving planet or short time series, PCA is expected to be less effective as the planet signal becomes more degenerate with these other effects. We discuss the implications of this for our own observations further in section \ref{sec:detection}.

\subsection{Model planet spectrum}\label{sec:Templates}

We generated model planet spectra using the Python package petitRADTRANS \citep{molliere_petitradtrans_2019}. We used the line-by-line mode of the package to generate a high-resolution (R=$10^6$) emission spectrum of a temperature-inverted UHJ atmosphere including only CO emission lines. We modeled WASP-33 b's thermally inverted dayside atmosphere using a two-point temperature profile, with a bottom temperature of 3000 K at P$_{bottom}$=10$^{-1}$ bar and top temperature of 4000 K at P$_{top}$=10$^{-3}$ bar. This profile is consistent with the constraints obtained from CO lines by \cite{darveau-bernier_caracterisation_2023} on the archival SPIRou observations, and those obtained on separate observations with KPIC by \cite{finnerty_keckkpic_2023}. Based on the retrieval by \cite{finnerty_keckkpic_2023}, we adopted a constant CO mass mixing ratio of log$_{10}$CO = -1.1, which is also consistent with the \cite{darveau-bernier_caracterisation_2023} joint constraints from SPIRou and HST. We used the Rayleigh scattering cross-sections computed for for an atmosphere dominated by H$_2$ and He, and collision induced absorption continuum opacities for H$_2$-H$_2$ and H$_2$-He pairs. We used the CO line opacities for the main isotopologues up to 5000 K, based on the line list from \cite{li_rovibrational_2015}. We generated the spectrum over the wavelength range 2.1--2.5 $\mu$m, covering the CO ro-vibrational band-head in the K band. As shown in \cite{finnerty_keckkpic_2023}, the H- opacity varies little over one spectral order in this wavelength range, so we chose to ignore it. For completeness, the spectrum generated from this model atmosphere is reproduced in Appendix \ref{app:Figs}, Figure \ref{fig:CO_spectrum}.

\subsection{Cross--correlation}

After applying the data processing described above, the planetary spectrum remains buried under the residual noise. As such, the spectra need to be effectively filtered in order to reveal the planet's trace. As typically done in the literature, we do this by cross-correlating the processed observational data with a model planetary spectrum. We used the same spectrum as in the signal injection for our planet template (see section \ref{sec:Templates}). We discuss the potential biases of this approach and compare its results against that of the real planet in section \ref{sec:mod_comp}.

As described in section \ref{sec:sig_inj}, this model was convolved to SPIRou's resolution and shifted to the planetary velocity for each frame in the sequence following equation \ref{eq:KP-Vsys}, for a range of $K_p$ and $V_{sys}$ values. Prior to performing the correlation, each sequence was degraded following \cite{gibson_relative_2022} to account for the deformations induced by PCA on the real signal (see \citealt{klein_atmospherix_2024}, section 3.3.3). We then computed the cross--correlation function,
\begin{equation}
    CCF = \sum_i \frac{f_im_i}{\sigma^2_i},
\end{equation}
where $f_i$, $\sigma_i$, and $m_i$ are the observed flux, its uncertainty, and the model value at each pixel $i$. Following \cite{boucher_characterizing_2021} and \cite{klein_atmospherix_2024}, we define $\sigma_i$ based on the standard deviation of each pixel at given wavelength $\lambda$, weighted by the S/N of each spectrum at a given time $t$:
\begin{equation}
    \sigma_i^2 = \sigma^2 (t,\lambda) = \frac{\Sigma_t (f(t,\lambda) - \overline{f(\lambda)} )^2}{N_{spectra}} \frac{\overline{SNR^2}}{SNR(t)^2}.
\end{equation}
Here $N_{spectra}$ is the number of spectra in the sequence and the bar denotes time-averaged values. We computed the correlation for models on a grid of $K_p$ and $V_{sys}$, spanning the expected/injected values of $K_p \pm 150$ km/s and $V_{sys} \pm 100$ km/s, in steps of 2 and 1 km/s respectively. The resulting correlation maps were converted into significance maps by dividing by the standard deviation away from the expected planet position, taken to be anywhere further than $50$ km/s away in both K$\rm_p$ and V$\rm_{sys}$.

\subsection{Log-likelihood mapping}\label{sec:loglike}

While the cross--correlation method described in the previous section is useful to efficiently detect the planetary signal and estimate its significance level, a more robust framework is required to compare different models and obtain realistic error bars (e.g: \citealt{brogi_retrieving_2019}). To do this, we used the likelihood framework of \cite{gibson_detection_2020}, which defines the likelihood $\mathcal{L}$ as
\begin{equation}
    \mathcal{L}=\prod_i \frac{1}{\sqrt{2 \pi (\beta\sigma_i)^2}} \exp \left\{-\frac{\left[f_i- \alpha m_i\right]^2}{2(\beta \sigma_i)^2}\right\},
\end{equation}
where $\alpha$ and $\beta$ are scaling parameters for the model and noise, respectively. Since the likelihood is much more sensitive to the line depths and shapes than the CCF, we need to take rotational broadening into account in the models in order to accurately fit the data with this framework. In our ingress--egress observations, this broadening depends both on the rotation velocity of the planet and its unconcealed portion, thus varying over time as the planetary disk is hidden or revealed. To implement this time-variable broadening more efficiently in model fits, we pre-computed a grid of broadened templates over a range of radial and rotational velocities, which we then linearly interpolate over for each model and phase according to the RV anomalies computed from \starry as described in section \ref{sec:sig_inj}. The resulting templates were finally scaled with the \starry outputs as described in section \ref{sec:sig_inj}, mean subtracted, and degraded to account for the effects of data processing with PCA -- if used -- before the likelihood was computed.

In our implementation of the likelihood framework, we fit for the model scaling $\alpha$, and then scale $\beta$ such that the minimum reduced $\chi^2_r$ is equal to 1 for the best-fitting model, as suggested in section 3.3 of \cite{gibson_detection_2020}. In practice, this means that we compute the $\chi^2$ as
\begin{equation}\label{eq:chi2}
    \chi^2 = \frac{1}{\beta^2}\left[\sum \frac{f_i^2}{\sigma_i^2} + \alpha^2 \sum \frac{m_i^2}{\sigma_i^2} - 2\alpha \sum \frac{f_i m_i}{\sigma_i^2}\right],
\end{equation}
where the first term is constant for a given dataset, and the second and third terms need to be computed for each model. Once these terms are computed for a given model, the $\chi^2$ is computed as a function of $\alpha$ following equation \ref{eq:chi2}, where $\alpha$ is optimised to minimise the $\chi^2$ over the range (0,2). Finally, the $\chi^2$ is mapped to the likelihood according to
\begin{equation}
    \ln \mathcal{L}=-\frac{N}{2} \ln \frac{\chi^2}{N},
\end{equation}

where $N$ is the number of data points. We computed the likelihood on the same grid as the CCF for K$_p$ and V$_{sys}$ which was extended for rotation velocity V$_{rot}$ and day--night contrast Y$_{l,m}(1,0)$, over a grid spanning $-50$ to $50$ km/s in steps of $1$ km/s and 0 to 1 in steps of 0.5 respectively. We computed error bars on the final likelihood maps using the $\Delta\chi^2$, where the uncertainty at N-$\sigma$ is given by the region over which $\chi^2\leq\chi^2_{min}+N^2$, where $\chi^2_{min}$ is the minimal value of the $\chi^2$, and $\beta$ is set such that the minimal reduced $\chi^2_r$ is 1.

\subsection{Injection-recovery tests}\label{sec:sig_inj}

To test detection limits and constraints that could be obtained with further data, we created a mock dataset by injecting a synthetic planet model into our data. To do so, we started by converting the spectrum generated in section \ref{sec:Templates} into a flux ratio between planet and star $F_p/F_*$, by dividing it with a black-body at WASP-33's temperature, and the ratio of star to planet radii squared.  For each frame in the injection sequence, this spectrum was weighted by the flux curve obtained from \starry in section \ref{sec:starry}, accounting for the change in line strength due to planet phase/visible portion. Next, a phase-dependant broadening was applied to account for the changing line shape during occultation. This was done using a rotational broadening kernel of half width equivalent to the injected planet's rotation velocity minus the RV offset calculated in section \ref{sec:starry}. While this doesn't fully capture the potential wing effects and/or asymmetries of a real line profile, it provides a good first order estimation of the broadening to which SPIRou is sensitive to. Finally, we simulated the instrumental broadening by convolving with a Gaussian of half width 4.29 km/s, corresponding to one SPIRou resolution element (or $\sim$two detector pixels).

The obtained sequence of planet spectra (in $F_p/F_*$) was Doppler-shifted according to the following equation:
\begin{equation}\label{eq:KP-Vsys}
    RV_p(\phi) = K_p \sin(2\pi \phi) + V_{sys} + \Delta RV_p(\phi).
\end{equation}
Here $K_p$ is the orbital semi amplitude, $V_{sys}$ the systemic velocity offset and $\Delta RV_p$ is the RV offset calculated in section \ref{sec:starry}. We set the K$_p$ and V$_{sys}$ of the synthetic sequence to 231 km/s and 150 km/s respectively, thus avoiding the real planet's position (K$_p$-231 km/s,V$_{sys}$-0 km/s), while keeping the same orbital amplitude. For the in-occultation data, further injections were made by increasing the V$_{sys}$ offset in increments of 50 km/s ($\sim$ 20 SPIRou pixels) to simulate different noise realisations from up to 40 observations (i.e. 20 sets of ingress--egress pairs). We also created some injection datasets with an amplified model, where we take an amplification factor $\sqrt{X}$ to represent X observations. 

\section{Results}\label{sec:Results}

In total, we have ten ingress and eight egress observations: seven of each from programme 24B, three partial ingresses from programme 23B, and one egress in programme 19B, providing a total symmetrical coverage of eight ingresses and egresses. In this section, we first discuss the results on this eclipse data alone, before presenting combined results including the full 19B programme data, and finally compare these to injection-recovery tests.

\subsection{Detecting CO in WASP-33b eclipses}\label{sec:detection}

To detect CO in WASP-33b, we used the model spectrum described in section \ref{sec:sig_inj}, the same one that we use for injection recovery tests. We validate this model on the archival dayside data from programme 19B, where it gives an S/N=5 detection at K$_p$=227$\pm$5 km/s  and V$_{sys}$=0$\pm3$ km/s consistent with the results from \cite{darveau-bernier_caracterisation_2023}. Within our dataset of eight eclipses of WASP-33 b, we detect a CO emission signal in cross--correlation at an S/N of 5, at K$_p$=247$^{+10}_{-18}$ km/s and V$_{sys}$=4$\pm$4 km/s. 

\begin{figure}[h!]
    \centering
    \includegraphics[width=0.96\linewidth]{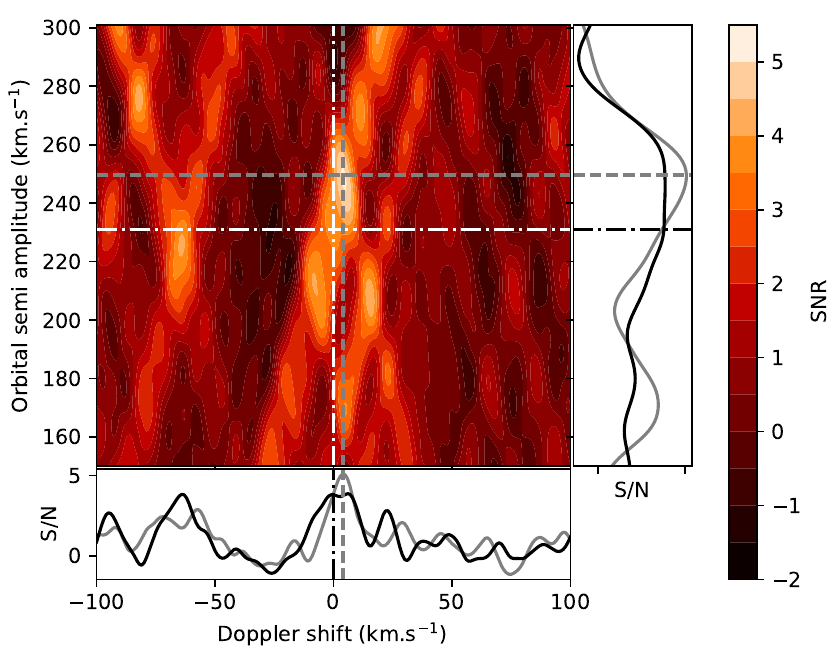}
    \caption{CCF map for eight ingress--egress pairs observed with SPIRou, showing a detection of CO in emission at S/N=5 at K$\rm_p$=247$^{+10}_{-18}$ km/s and V$\rm_{sys}$=4$\pm$4 km/s. Cross-sections at the peak S/N (grey dashed lines) and expected planet position (white dash-dotted lines) are shown in grey and black in the side and bottom panels for K$\rm_p$ and V$\rm_{sys}$, respectively.}
    \label{fig:KP-Vsys_Occs}
\end{figure}

As seen in figure \ref{fig:KP-Vsys_Occs}, this detection presents a double-peaked structure, with a maximum that is quite offset in K$\rm_p$ compared to the expected velocity. This offset, which may seem quite large compared to other emission datasets, can nevertheless be explained when considering the eclipse mapping signal (see Figure \ref{fig:starry_outputs}). As shown in Figure \ref{fig:eclipse_model_only} of the appendix, the velocity offsets induced during eclipse ingress--egress result in large velocity shifts in K$\rm_p$/V$\rm_{sys}$ maps. Specifically, in the case of WASP-33 b,\footnote{assuming synchronous rotation} the offsets induced over the short time frames of ingress--egress almost compensate completely for the planet's orbital motion. In consequence, the ingress and egress trails are shifted away from each other, such that their intersection peaks at a higher orbital semi-amplitude (see Appendix \ref{app:Model_CCF}).

\begin{figure}[h]
    \centering
    \includegraphics[width=0.96\linewidth]{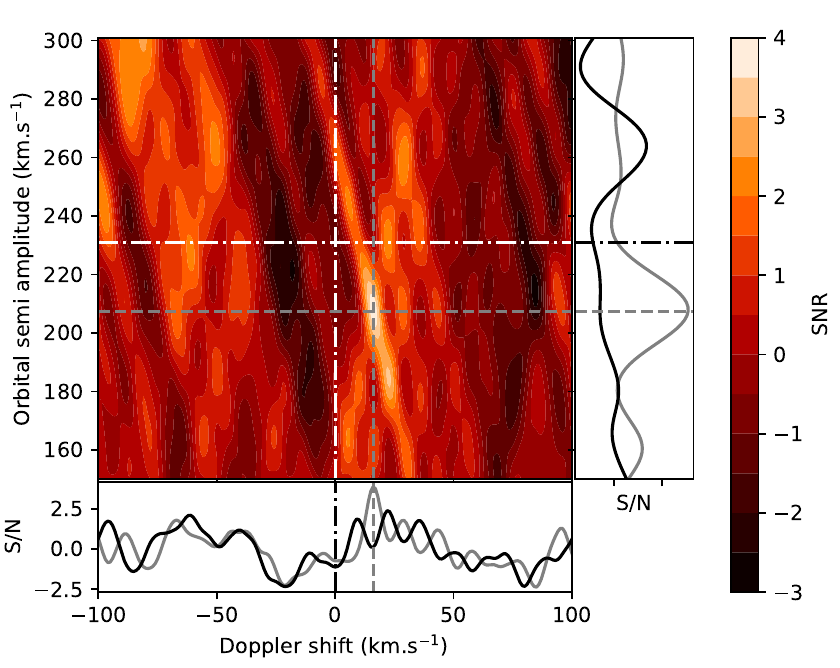}
    \caption{Same as Figure \ref{fig:KP-Vsys_Occs}, but considering only the frames in eclipse ingress--egress. The detection strength is lower, as the planetary contribution is reduced while it is being occulted, and the maximum is shifted to a lower orbital velocity along the diagonal corresponding to the contribution from ingress frames.}
    \label{fig:Kp-Vsys_In_eclipse}
\end{figure}

To confirm whether this shifted signal is indeed present during ingress--egress, we perform the CCF on the in-ingress--egress frames only. The results, shown in figure \ref{fig:Kp-Vsys_In_eclipse}, reveal a similar double-peaked structure, albeit with a peak that is shifted downwards along the trail corresponding to the ingress frames. The S/N is decreased to 3.8 at the peak, which is unsurprising as we are using less data, and the planetary contribution is reduced during the ingress--egress frames due to the occultation by the star. Similarly, while the nature of the detection remains tentative, the large velocity offset can be explained quite naturally if one side of the eclipse is contributing more than the other. As seen in Figure \ref{fig:eclipse_model_only} of the appendix, the eclipse mapping signal results in maxima being shifted to much lower orbital semi-amplitudes and larger positive (/negative) systematic velocity shifts when fitting on ingress (/egress) data alone. 

Such an imbalanced contribution could be expected in the case of a shifted hotspot/hemispheric asymmetry, however based on previous observations and simulations of WASP-33 b, this would be expected to lead rather to an asymmetry favouring the egress trail (see section \ref{sec:Hotspots}). Another explanation for this effect could be our eclipse timings being slightly offset. We tested using different ephemerides for the planet, namely those of \cite{zhang_phase_2018} and \cite{ivshina_tess_2022}, but found no difference in the position of the peaks whether including out of ingress--egress frames or not. We thus settled on the most recent ephemeris from \cite{smith_cheops_2025} leveraging TESS and CHEOPS data, and which resulted in very slightly higher S/N values.\footnote{This is likely due to the weighting of frames by the eclipse window in the CCF, and did not affect the error bars on the velocity.} While our detection strength is not strong enough to motivate interpreting this imbalance in ingress--egress frames physically, we take these tentative detections as evidence that the eclipse mapping signal is stacking coherently between events, and thus that stacking eclipses can be a viable strategy to increase the Doppler mapping sensitivity.

\subsection{Joint rotation fit}

We fit for the rotation velocity of WASP-33 b using the log-likelihood framework, which enables more robust model inter-comparison than the CCF. As discussed in section \ref{sec:PCA}, despite using the projector from \cite{gibson_relative_2022}, the PCA may significantly alter the planetary signal in short time series, leading to narrower lines and preferentially erasing the eclipse mapping signal. In our injection-recovery tests, this resulted in model fits consistently recovering models with no rotation velocity, reflecting an artificial preference for less broadened lines induced by the effects PCA. As such, we refrained from using PCA for the observations from programmes 23B and 24B, and maintain its use only for the observations from programme 19B, which do not suffer from this short time series issue. 

Due to the lack of PCA and short baseline, fits on the eclipse data alone do not yield any strong constraints yet, despite a peak which can be seen slightly below the noise level at the expected velocity of the planet (based on the 19B data). Meaningful constraints can however be obtained by combining the eclipse datasets with the archival dayside phase observations and comparing fits with and without including occultations. The results, summarised in Table \ref{tab:results_comp}, show that the constraints on K$\rm_p$ and V$\rm_{sys}$ remain largely unchanged when including eclipses, as they are already well constrained thanks to the larger velocity excursions present in the longer dayside observations. In contrast, the uncertainty on the rotation velocity decreases significantly when eclipses are included. While the inferred rotation remains consistent with synchronous rotation at the 1$\rm\sigma$ level, the double-peaked structure seen in the phase data fits somewhat disappears when including eclipses, leaving place to a single peak centred around a positive (prograde) rotation velocity. This behaviour suggests that, while the constraint on rotation from our phase data comes from broadening alone (hence the double-peak), the eclipses may be stacking coherently to provide a directional signal favouring prograde rotation. We stress however that the current precision remains insufficient to claim a robust detection of the velocity offsets, and further observations will be required to unambiguously detect the eclipse mapping signal.

\begin{table}[h!]
\caption{Best-fit parameters and 1$\sigma$ errors from the log-likelihood fits of WASP-33 b's CO emission.}\label{tab:results_comp}
\centering
\begin{tabular}{ccc}
\hline \hline
Parameter & Daysides Only    & Daysides+Eclipses \\ \hline
K$_p$        & 228 ($\pm$3) km/s & 228 ($\pm$2) km/s  \\
V$_{sys}$     & 0 ($\pm$1) km/s    & 0 ($\pm$1) km/s    \\
V$_{rot}$      & -4 (+10 -4) km/s  & 2 ($\pm$5) km/s      \\
day--night Contrast     &    --             &       --               \\ \hline
\end{tabular}
\end{table}

\subsection{Injection recovery: Towards a rotation measurement}

\begin{figure}[h!]
    \centering
    \includegraphics[width=0.98\linewidth]{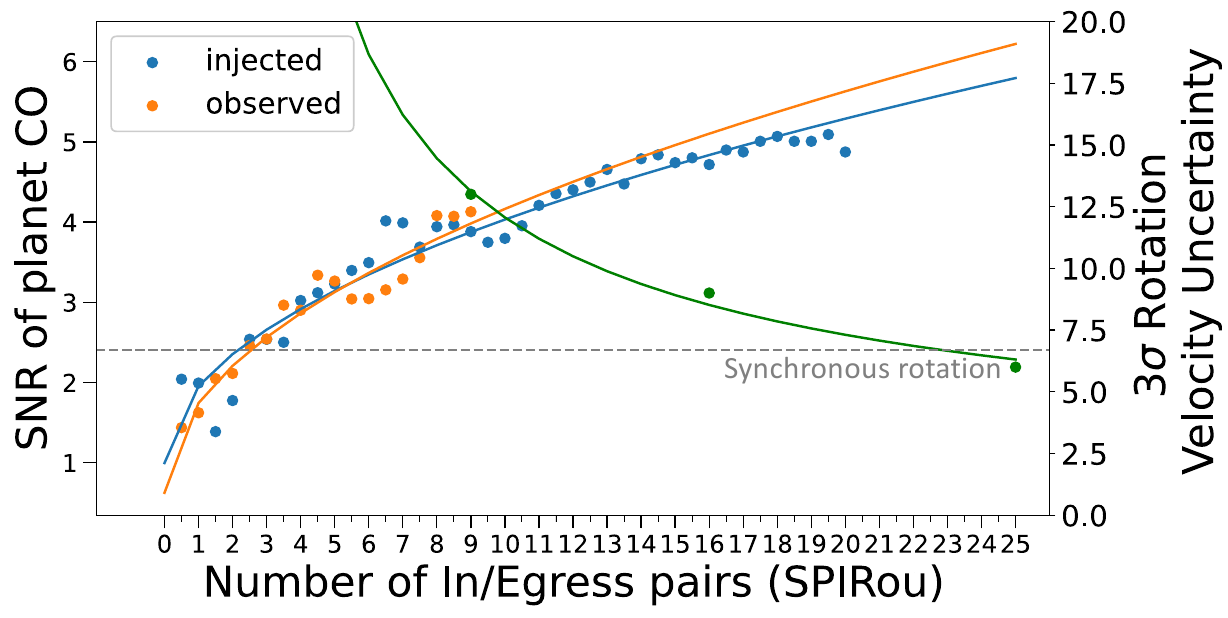}
    \caption{Scaling of the CO detection from eclipse data in injected vs real SPIRou observations of WASP-33b, and the corresponding 3$\rm\sigma$ uncertainties on the planet's rotation velocity in injection-recovery tests (green). Indicative square root trends are plotted in line with the expectations from photon-noise statistics. The precision to detect synchronous rotation at 3$\rm\sigma$ is reached between 20 and 25 eclipses in injection-recovery tests.}
    \label{fig:SN_vrot}
\end{figure}

We now perform the same analysis on a set of injected eclipse observations (see section \ref{sec:sig_inj}) in order to reach realistic detection limits for what could be achieved with further data from SPIRou. As shown in section \ref{sec:detection}, the CCF maps behave qualitatively the same between our observed eclipses and injection-recovery tests. This is shown more quantitatively in figure \ref{fig:SN_vrot}, which compares the maximum S/N of the CCF at the planet position in relation to the number of eclipses in real vs injected observations. As can be seen on the figure, the S/N follows a comparable evolution in both datasets, broadly consistent with the square-root trend lines plotted for reference.\footnote{A square-root dependence is the naïve expectation from photon-noise scaling, although it may break down if a noise floor is approached after a certain number of observations.} Extrapolating this trend suggests that an S/N above 5 can be reached after approximately 15 to 20 eclipses. We caution, however, that this estimate still relies on the use of two principal components in the de-trending, which, as demonstrated in our injection–recovery tests, acts to preferentially remove the rotationally induced velocity offset.

To test the performance of Doppler mapping we thus omit PCA entirely, meaning that the results presented here may ultimately be improved if alternative de-trending methods can be implemented that preserve the planetary signal. We applied the log-likelihood framework to injected datasets using the amplified model (for computational efficiency), testing injections corresponding to 9, 16, and 25 ingress--egress pairs. The resulting 3$\sigma$ uncertainties on the recovered rotation velocity are summarised in green in Figure \ref{fig:SN_vrot}, along with a trendline in 1/$\sqrt{X}$, where X is the number of stacked observations. As can be seen from the trendline, the precision should reach below the synchronous-rotation somewhere between 20 and 25 eclipses, indicating that another 10 to 15 eclipses will be required with SPIRou to obtain meaningful rotational constraints from eclipse-only data. 

The full posterior distributions for the 25-eclipse case are shown in blue in Figure \ref{fig:Corner_full}, where they are compared to the constraints obtained when injecting a single set of out-of-eclipse phases into the 19B programme data (in orange). In the phase-only case, the longer temporal baseline improves the constraint on the planetary orbital velocity amplitude, but provides little information on the rotation rate. This leads to the characteristic double-peaked posterior in rotation velocity, a degeneracy that arises because the dayside line broadening is symmetric with respect to prograde and retrograde rotation. In contrast, the 25-eclipse scenario yields a broader constraint on the orbital velocity amplitude but a single, well-defined peak in rotation velocity. The precision achieved is sufficient to distinguish synchronous rotation from no rotation at the 3$\sigma$ level, thanks to the additional constraint provided by the rotational velocity offset captured during ingress and egress. This result implies that a total of $\sim$25 eclipses should enable a 3$\sigma$ detection of the eclipse mapping signal for a synchronously rotating WASP-33 b, with the rotation rate measured precisely enough to exclude a non-rotating solution.

We also find that the fit remains largely insensitive to the day–night contrast ratio. This behaviour is not surprising considering our very short phase coverage and the loss of continuum flux at high spectral resolution. In future, combining these eclipse constraints with longer HRS phase curves and/or space-based observations could help to break this degeneracy; although further work is required to determine how and whether constraints from LRS and HRS can be meaningfully combined. More generally, our results demonstrate that combining phase-curve and eclipse spectroscopy at high resolution provides a significantly more powerful diagnostic than either dataset independently, as each observation type constrains complementary aspects of the atmospheric velocity field. With sufficiently high S/N data, exploring alternative parametrisations of the day--night contrast, for example explicitly capturing how the hotspots impact the line shapes, may further improve constraints on the dayside temperature profiles from this method.

\section{Discussion}\label{sec:Discussion}

\subsection{Robustness of the injections}\label{sec:mod_comp}

We assess the realism of our injection–recovery framework by directly comparing the S/N values obtained from observed and injected datasets in the dayside observations. We find good agreement between the two, with a S/N of 4.8 recovered in the injection tests compared to a S/N of 5.0 in the real data (see Appendix \ref{app:Figs}, Figure \ref{fig:KP-Vsys}). This consistency is also reflected in the model fits themselves: the best-fitting scaling parameter ($\alpha$=0.8) is close to unity, indicating that the real signal's amplitude closely matches that of our atmospheric model. The small difference is likely attributable to the pre-processing steps other than PCA, which are not reproduced on the model in our current implementation. Despite this agreement, the injected model remains a simplified representation of WASP-33 b's atmosphere, and may overestimate the detectable signal during eclipse phases. Our injections rely on several assumptions, specifically, that the emission contribution is distributed smoothly throughout the dayside in a way which can be described by low spherical harmonic degree orders, and that the line shape can be adequately captured by a simple rotational broadening kernel. While these are likely adequate approximations at first order as shown by the matching S/N trends (Figure \ref{fig:SN_vrot}), more complex spatial distributions and/or wind-driven velocity fields may impact the shape and amplitude of the eclipse mapping signal in non-trivial ways. Further observations will be required to obtain a robust detection of this signal and assess the potential impact of these parameters on its detectability.

\subsection{Impact of PCA on model fits}\label{sec:PCA}

In light of the relatively low S/N of our detections, an alternative interpretation for the positive offset and/or double peak structure seen in figure \ref{fig:KP-Vsys_Occs} could be as an artifact due to the use of PCA. As hinted at in section \ref{sec:data_proc}, our very short time series mean that the planet signal moves only very little ($\rm\sim$15 km/s or $\rm\sim$7 pixels over 6 exposures). Furthermore, the rotational broadening of the planet (V$\rm_{syn}$=6.7 km/s for a synchronous rotation) acts to spread the spectral lines over 6 pixels (3 pixels either side).  As such, the planetary signal in a given pixel is largely time-correlated, meaning a significant portion of it could be removed by the PCA de-trending, thus biasing the final detection to higher orbital amplitudes and lower rotational broadening. 

This can be seen in Figure \ref{fig:CCF_rest_frame} of the appendix, showing the planetary rest frame CCF for an injection test with an amplified planet signal, where the trace of the planet is clearly seen in the CCF before applying PCA. When applying PCA, fixed velocity residuals outside the planetary trail are visibly removed, resulting in a stronger detection. However, the PCA also seems to degrade the underlying planetary signal along this fixed velocity axis, such that the final peak is much narrower and the planetary trail much less visible in the planet rest-frame CCF, despite having corrected the model template for the effects of PCA with the projector from \cite{gibson_relative_2022}. This is all the more prominent for the in-ingress--egress spectra, where the velocity offset due to the planet's rotation acts to reduce its frame-to-frame velocity excursion, leading to PCA preferentially removing the part of the signal which we are trying to fit for. As such, while beneficial for improving detection limits for a weak signal, PCA should be avoided in model fits exploiting these short eclipse observations until these effects can be understood more fully and/or mitigated by alternative de-trending methods. In the meantime, our recommendation for similar programmes would be to favour slightly longer baselines (e.g. $\sim$1-2 hours vs our $\sim$45 minute time series) with the planet in view, such that PCA may be used without introducing such strong biases.

\subsection{Hotspots and hotspot offsets}\label{sec:Hotspots}

Our use of a low-order spherical harmonic intensity map of the planet's CO emission (section \ref{sec:starry}) captures structure of a UHJ to first order, with a hot, highly contributing dayside and a cooler, poorly contributing nightside. Dayside hotspots, if present, should act to steepen the contribution slope, which we fit as our day--night contrast. Interestingly however, both injection tests and fits on the observed data reveal very weak sensitivity to this parameter, meaning that models with homogeneous emission across the entire planet are not discarded by our fits (see Figure \ref{fig:Corner_full}). As shown in Figure \ref{fig:starry_outputs}, the main impacts of a stronger hotspot outside eclipse are larger velocity offsets and a slight decrease in line contrast, while its impacts are almost indistinguishable during eclipse. It is therefore not particularly surprising that we do not strongly prefer models including a hotspot, as our phase coverage is not high enough to expect velocity shifts above a few km/s or line contrast changes above $\sim$10\%. 

For completeness, we did try fitting a hotspot-driven phase offset on the archival dayside data, which has larger pre and post eclipse coverage. In the case of WASP-33b, hotspot-driven phase offsets have already been proposed to explain features detected in HRS, for example by \cite{herman_dayside_2022}, who find a positive phase offset in their iron detection with ESPaDOnS, which they interpret as indicative of a westward hotspot offset of +22$\pm$12 degrees. \cite{van_sluijs_carbon_2023} also found a positive phase offset based on a higher scaling factor post-eclipse in their CO detection, however they interpret this as indicative of an eastward hotspot offset based on GCM simulations (see section \ref{sec:Doppler}). Consistently with these studies, we also see a preference for higher scaling factors post eclipse when fitting phases independently, with an optimal $\rm\alpha$ value of 0.46 and 1.44 pre- and post-eclipse respectively. Curiously however, this translates into only very weak constraints on the hotspot offset in joint fits using the scaling and velocity offsets from our \starry simulations, and including the possibility of such an offset does not change the optimal value of the global $\rm\alpha$ even in separate fits. As such, we conclude that our data do not have the required sensitivity, either in phase coverage and/or S/N, to confidently recover hotspot properties, and that the large difference in scaling factors is more likely due to the different observing conditions between the two nights. Further observations with SPIRou will be required to reach higher detection S/N values in eclipse and longer phase coverage, which will hopefully enable such fits to yield useful constraints.

\subsection{Physical interpretation of Doppler maps}\label{sec:Doppler}

In contrast to their low-resolution counterparts, high-resolution phase and eclipse Doppler maps cannot be interpreted as direct temperature maps of the planetary surface. The resolving power of HRS allows individual atoms/molecules to be isolated, so that Doppler maps trace the spatial distribution of specific species rather than the bulk atmosphere. Different species are thus expected to produce distinct maps, reflecting variations in chemical abundance, condensation, dissociation, and vertical mixing across the planet \citep{wardenier_pre-transit_2025}, and each species must be considered independently. Moreover, Doppler maps are not sensitive to absolute temperature but to the vertical temperature gradient, which controls the depth and contrast of spectral lines. As a result, apparent hotspots at high spectral resolution do not necessarily coincide with the hottest regions of the atmosphere. GCMs with eastward-shifted hotspots often predict steeper temperature gradients on the western dayside, leading to stronger post-eclipse line contrasts, as has already been observed for WASP-33 b \citep{herman_dayside_2022,van_sluijs_carbon_2023}.

In addition, Doppler maps are intrinsically sensitive to atmospheric velocities through both line-of-sight shifts and rotational and dynamical broadening. Variations in the vertical and horizontal distribution of molecules can result in line profiles that differ from species to species. Such behaviour has been clearly demonstrated in transit spectroscopy, where different molecules exhibit distinct Doppler shifts and line shapes linked to their vertical distribution and morning–evening asymmetries (e.g. \citealt{flowers_high-resolution_2019,nortmann_crires_2025,hood_atmospherix_2024,seidel_vertical_2025}). Similar effects have been reported in emission studies, where different species show offsets in K$\rm_p$ and V$\rm_{sys}$ suggestive of spatially varying winds and abundances
(e.g. \citealt{cont_detection_2021,bazinet_quantifying_2025}). The latitudinal distribution of absorbers can further modify rotational broadening \citep{charnay_latitudinal_2025}, and complex wind geometries can distort line shapes in non-trivial ways \citep{lesjak_retrieving_2025}. While these higher-order effects remain difficult to disentangle with the S/N values afforded by current facilities, they will become crucial to consider with the next-generation instruments such as ANDES \citep{palle_ground-breaking_2025}.

\subsection{Prospects for ELT}

\begin{figure}[h!]
    \centering
    \includegraphics[width=\linewidth]{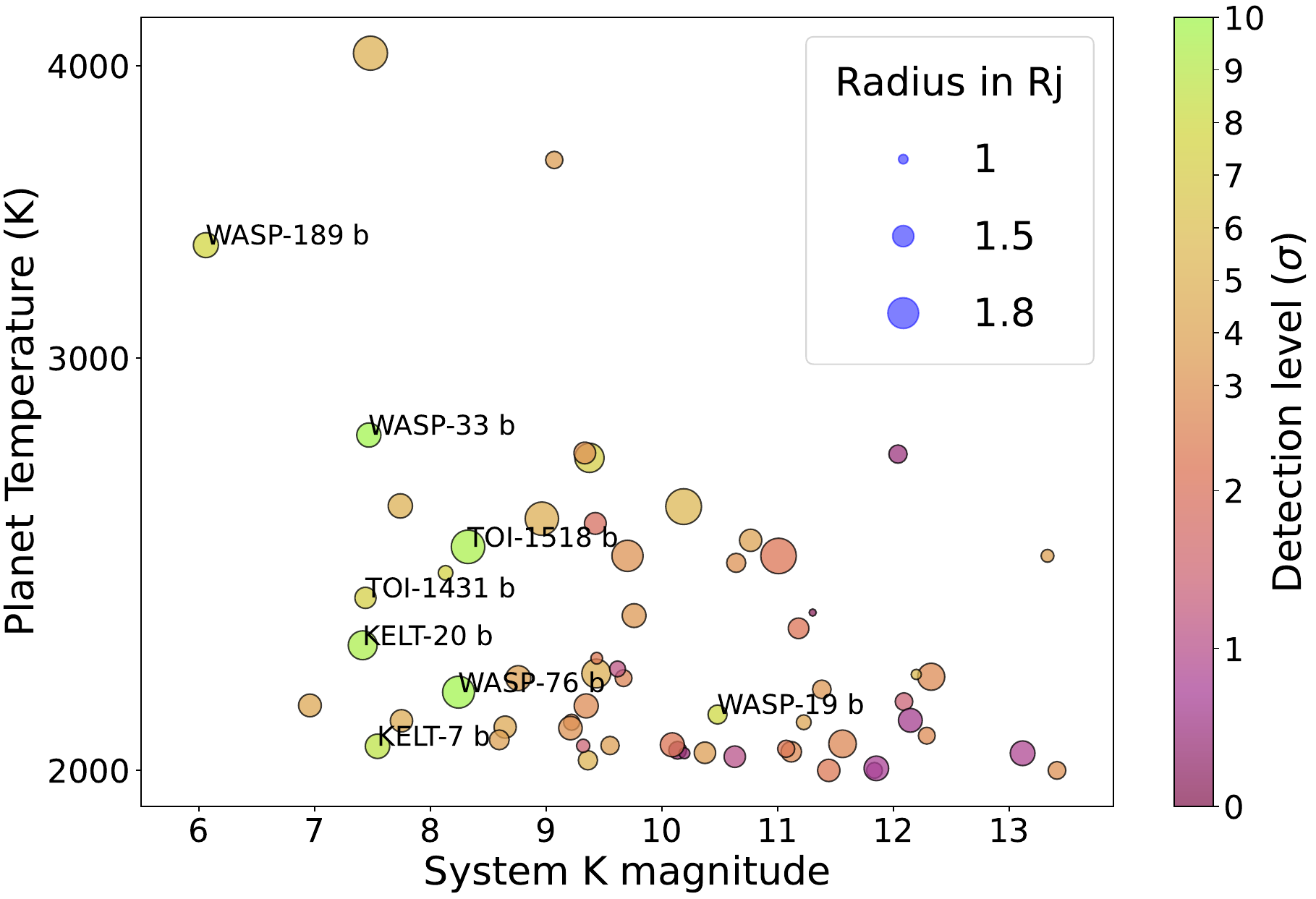}
    \caption{Photon-limited detection S/N of the radial velocity offset induced by tidally locked rotation rates for a population of UHJs, measurable from CO lines in the K band in one occultation with ELT/ANDES.}
    \label{fig:SNR_RV_ANDES}
\end{figure}

While Doppler eclipse mapping is likely beyond the reach of current facilities for most targets, the advent of the Extremely Large Telescope (ELT) is expected to transform this observational landscape. The substantially increased collecting area of 30–40 m class telescopes is critical for achieving the high signal-to-noise ratios required over the short ingress and egress timescales that encode the eclipse mapping signal. In this context, the ELT/ANDES instrument \citep{marconi_andes_2022}, with its high spectral resolution and broad wavelength coverage, is particularly well suited to Doppler eclipse mapping. Figure \ref{fig:SNR_RV_ANDES} presents the photon-noise–limited detection S/N values for the eclipse mapping signal achievable with ANDES in a single occultation across the current population of known ultra-hot Jupiters. For approximately 10–20 systems, a single eclipse would already suffice to yield a strong detection, while stacking multiple occultations would extend this capability towards cooler hot Jupiters. We note, however, that these estimates are based purely on the raw collecting power and expected throughput of the instrument and therefore represent optimistic limits. Our SPIRou simulations indicate that, in practice, roughly twice as many observations as predicted by photon-noise considerations alone may be required to achieve a robust detection. How this scaling translates to the ELT regime remains uncertain, particularly given the reduced impact of thermal background and systematics expected for larger telescope apertures.

Although ANDES is still several years away, first-light ELT instruments may already offer promising opportunities for Doppler mapping of UHJs. Figure \ref{fig:SNR_RV_ELT} compares the photon-limited performance of MICADO and ANDES for detecting the eclipse mapping signal of CO in the K band. Despite MICADO’s lower spectral resolution (R=20,000), a small number of targets are accessible within a single occultation. In particular, WASP-19 b emerges as a compelling candidate whose atmospheric dynamics could be probed with early ELT observations. While lower spectral resolution limits the level of detail that can be recovered from individual line profiles, previous work has already shown that resolutions of R$\simeq$10,000 can already be sufficient to probe three-dimensional atmospheric effects in UHJs \citep{van_sluijs_carbon_2023}. In fact, eclipse observations can be especially valuable in this regime, providing additional spatial information that helps disentangle atmospheric dynamics. In this context, combining eclipses with longer dayside time series would be a powerful strategy to maximise the information content of moderate-resolution datasets. Overall, these prospects underscore the need to develop retrieval frameworks capable of accounting for three-dimensional atmospheric structure and velocity fields, and motivate more detailed end-to-end simulations to establish realistic detection limits ahead of the ELT coming online.

\section{Conclusion} \label{sec:concl}

In this article we have demonstrated the first data-informed exploration of eclipse mapping at high resolution, revealing its potential to resolve the rotation, winds, and spatial contribution of molecules. Based on an observation programme of eight eclipses of the UHJ WASP-33b with CFHT/SPIRou and a range of injection-recovery tests on the same data, we have shown the following:

1. We detect CO emission for the first time from eclipse ingress--egress only, validating that eclipses can be stacked coherently to increase the detection limits on a given system.

2. Our detection is shifted towards higher orbital semi-amplitudes, consistently with expectations from simulations, providing tentative evidence that we are seeing velocity offsets due to WASP-33 b's rotation during eclipse ingress--egress.

3. Our detection strength is scaling according to expectations from injection-recovery tests, implying a further 15 eclipses ($\sim$ 15h) with SPIRou should allow us to confidently recover a synchronously rotating WASP-33 b from the velocity offsets observed during eclipse ingress and egress.

4. Much like its low-resolution counterpart, eclipse mapping at high resolution is more powerful when combined with longer baseline phase curve data, motivating a push to obtain partial or even full phase curves covering longer baselines with HRS on promising targets such as WASP-33b.

5. Current data reduction procedures suffer similar pitfalls when dealing with short time series and slow-moving planets, motivating the development of more robust de-trending tools, which could be tested on such short time series of planets with favourable contrasts ahead of next-generation telescopes.

Finally, informed by these initial results with SPIRou, we have provided some perspectives for the potential of this method with the upcoming ELT, showing that Doppler eclipse mapping may be achievable on a few targets with MICADO,  and will be a very promising science case for ANDES.

\begin{acknowledgements} 
This work is supported by the French National Research Agency in the framework of the Investissements d’Avenir program (ANR-15-IDEX-02), through the funding of the “Origin of Life" project of the Grenoble-Alpes University. V. Yariv acknowledges funding from the physics doctoral school (ED-PHYS) of the Grenoble-Alpes University as well as through MITACS globalink graduate funds. F. Debras acknowledges funding from the French National Research Agency (ANR) project ExoATMO (ANR-25-CE49-6598). F.Debras would like to warmly thank the \textit{Action Thématique Physique Stellaire} (ATPS) and \textit{Action Thématique ExoSystèmes} (AT-EXOS) of CNRS/INSU co-funded by CEA and CNES for their financial support. A. Masson acknowledges support and funds from the grants nº CNS2023-144309 by the Spain Ministry of Science, Innovation and Universities MICIU/AEI/10.13039/501100011033. R.A. acknowledges the Swiss National Science Foundation (SNSF) support under the Post-Doc Mobility grant P500PT\_222212 and the support of the Institut Trottier de Recherche sur les Exoplanètes (IREx).

This study is based on data obtained at the CFHT, operated by the CNRC (Canada), INSU/CNRS (France), and the University of Hawaii. The authors wish to recognise and acknowledge the very significant cultural role and reverence that the summit of Maunakea has always had within the indigenous Hawaiian community. We are most fortunate to have the opportunity to conduct observations from this mountain.

This work made use of the python packages astropy \citep{astropy:2013, astropy:2018, astropy:2022}, numpy \citep{harris_array_2020}, scipy \citep{virtanen_scipy_2020}, \starry \citep{luger_starry_2019}, petitRADTRANS \citep{molliere_petitradtrans_2019}. This research has made use of the Astrophysics Data System, funded by NASA under Cooperative Agreement 80NSSC21M0056. 
\end{acknowledgements}

%
%

\bibliographystyle{aa} 
\bibliography{bibliography}

\begin{appendix}
    \section{Computing photon-limited uncertainties on planetary radial velocities} \label{app:photon}

Measurement of rotation rates/winds with the Doppler eclipse mapping technique requires achieving a good enough precision on the planetary radial velocity (pRV). In the case where it is limited by photon noise, this precision can straightforwardly be computed based on the model planet and/or star spectra in photon counts, providing an efficient way to compute detection limits for a wide range of targets and instruments. Here, we describe how we perform such estimations, expanding the formalism for stellar radial velocities described in \citet{bouchy_fundamental_2001} to planetary radial velocities measured through HRS. 

Following \citet{bouchy_fundamental_2001}, we consider a noise-free model planet spectrum $A_{0p}$, which are Doppler shifted at various epochs to become $A_{p}$. The intensity and wavelength at each pixel $i$ in $A_{0p}$ are noted respectively $A_{0p}(i)$ and $\lambda(i)$. The Doppler shift of spectrum $A_{p}$ measured at each pixel $i$ is then expressed in terms of the intensity change as
\begin{equation}
    \frac{\delta V_p(i)}{c} = \frac{A_p(i) - A_{0p}(i)}{\lambda(i)\left(\delta A_{0p}(i)/\delta \lambda(i)\right)} .
\end{equation}

The contributions from each pixel are summed over the entire spectral range according to an optimal weight $W(i)$. This is proportional to the inverse square of the dispersion in the velocity measured at the pixel $i$, given by
\begin{equation}
    \frac{\delta V_{pRMS}(i)}{c} = \frac{\left[A_p(i) - A_{0p}(i)\right]_{RMS}} {\lambda(i)\left(\delta A_{0p}(i)/\delta \lambda(i)\right)},
\end{equation}
and results in the optimum weighting function,
\begin{equation}\label{eq:weight}
    W(i) = \frac{1}{\left(\frac{\delta V_{pRMS}(i)}{c}\right)^2} = \frac{\lambda^2(i)\left(\delta A_{0p}(i)/\delta \lambda(i)\right)^2}{\left[A_p(i) - A_{0p}(i)\right]_{RMS}^2},
\end{equation}
where $\left[A_p(i) - A_{0p}(i)\right]_{RMS}$ designates the dispersion in intensity change measured at pixel $i$. In the case of an isolated star or planet,  this quantity is limited by the intensity of the model spectrum and expressed straightforwardly as $\left[A_p(i) - A_{0p}(i)\right]_{RMS} = \sqrt{A_{p}(i) + \sigma_D^2}$, where $\sigma_D^2$ is the contribution from the detector noise. In the case of a close-in planet however, the photon noise and thus dispersion of our intensity measurement are dominated instead by the stellar photon noise. The dispersion is thus  expressed as
\begin{equation}
    \left[A_p(i) - A_{0p}(i)\right]_{RMS} = \sqrt{A_*(i)+A_p(i)+\sigma_D^2},
\end{equation}
where $A_*$ is the intensity of a model stellar spectrum corresponding to the planet's host star. For a small Doppler shift, we can set $A_p = A_{0p}$ and write the weighting functions $W_p(i)$ and $W_*(i)$ in the planetary and stellar noise dominated cases respectively as
\begin{equation}
    W_p(i) = \frac{\lambda^2(i)\left(\delta A_{0p}(i)/\delta \lambda(i)\right)^2}{A_{0p}(i)+\sigma_D^2},
\end{equation}
\begin{equation}
    W_*(i) = \frac{\lambda^2(i)\left(\delta A_{0p}(i)/\delta \lambda(i)\right)^2}{A_*(i)+A_{0p}(i)+\sigma_D^2}.
\end{equation}
The first equation is equivalent to the weighting function $W(i)$ in equation 8 of \citet{bouchy_fundamental_2001}, applied to the planetary spectrum, while the second is a new weighting function accounting for the stellar noise, which is in our case the dominant form of noise. 

The Doppler shift measured over the entire spectral range is then given by the weighted sum:
\begin{equation}
    \frac{\delta V_p}{c} = \frac{\sum \frac{\delta V_p(i)}{c} W_*(i)}{\sum W_*(i)} = \frac{\sum
    (A_p(i)-A_{0p}(i))\left(\frac{W_*(i)}{A_*(i)+A_{0p}(i)+\sigma_D^2}\right)^{1/2}}{\sum W_*(i)}.
\end{equation}
Considering the stellar flux is orders of magnitude larger than the planet's and assuming negligible detector noise, we can express the new weighting function in terms of the former as
\begin{equation}
 W_*(i) = W_p(i) \frac{A_{0p}(i)}{A_*(i)}.
\end{equation}

Then finally, the uncertainty $\delta V_{RMS}$ in the radial velocity measurement is given from equation \ref{eq:weight} by
\begin{align}
\delta V_{pRMS} &= \frac{c}{\sqrt{\Sigma W_*(i)}}
                = \frac{c \sqrt{\Sigma A_*(i)}}{\sqrt{\sum W_p(i)} \sqrt{\sum A_{0p}(i)}} \\
\text{or } \Aboxed{\delta V_{pRMS} &= \frac{c \sqrt{\Sigma A_*(i)}}{Q \sum A_{0p}(i)}},
\end{align}
where $Q= \frac{\sqrt{\Sigma W_p(i)}}{\sqrt{\Sigma A_{0p}(i)}}$ corresponds to the flux-independent quality factor as defined in \citet{bouchy_fundamental_2001}, evaluated on the planetary spectrum.

\section{Instrument parameters used in photon noise computations}

In table \ref{tab:Instruments}, we list the values of instrumental resolution, total throughput and telescope diameters that were used to obtain the photon noise estimates shown in figures \ref{fig:SNR_RV}, \ref{fig:SNR_RV_ANDES}, and \ref{fig:SNR_RV_ELT}.

\begin{table}[h!]
    \caption{Parameters used in photon noise computations for each of the instruments considered in this work.}
    \label{tab:Instruments}
    \resizebox{\linewidth}{!}{\begin{tabular}{lcccc}
    \hline
       Instrument & \begin{tabular}[c]{@{}c@{}} Telescope \\ Diameter \end{tabular} & \begin{tabular}[c]{@{}c@{}} Overall \\ Throughput \end{tabular} & Resolution & Ref. \\ \hline
        SPIRou & 3.6 m & 0.12 & 70 000 & [a] \\ 
        MICADO & 39.3 m & 0.15* & 20 000 & [b] \\
        ANDES  & 39.3 m & 0.06* & 100 000 & [c] \\ \hline
    \end{tabular}}
    \begin{tabular}{p{\linewidth}}
        *While no official total throughput estimates could be found for these ELT instruments, these values were taken as reasonable guesses based on existing HRS instruments on 8m class telescopes\\
        {[}a{]} \cite{donati_spirou_2020} \\ 
        {[}b{]} \cite{sturm_micado_2024} \\ 
        {[}c{]} \cite{marconi_andes_2022} \\
    \end{tabular}
\end{table}

\section{Summary of observations}\label{app:Obs}

In table \ref{tab:obs}, we show a summary of all the SPIRou observations obtained and used in this work. The table also presents the airmasses, S/N values and phase-coverage of each dataset, and individual exposure times for each time series.

\begin{table*}[h!]\label{tab:obs}
\centering
\caption{Dayside observations of WASP-33 with SPIRou used in this work.}
\begin{tabular}{cccccccccc}
\hline
\multicolumn{1}{l}{Observation} & \multicolumn{1}{l}{UT Date} & \multicolumn{1}{l}{Exposure Time} & \multicolumn{1}{l}{Seeing} & \multicolumn{1}{l}{S/N \textdagger} & \multicolumn{1}{l}{In Eclipse} & \multicolumn{1}{l}{Ingress} & \multicolumn{1}{l}{Egress} & \multicolumn{1}{l}{Pre/Post Eclipse} & \multicolumn{1}{l}{Total} \\ \hline
1* & 2019-10-07 & 267s & 0.49-0.66 '' & 91 & 10 & - & 4 & 0/24 & 38 \\
2* & 2019-11-04 & 267s & 0.76-0.91 '' & 80 & 0 & 2 & - & 36/0 & 38 \\
3 - 4 & 2023-12-10 & 245s & 0.54-0.65 '' & 92 & 0+8 & 3 & 0 & 5/0 & 16 \\
5 - 6 & 2023-12-16  & 245s & 0.48-0.57 '' & 88 & 0+8 & 3 & 0 & 5/0 & 16 \\
7 - 8 & 2023-12-27 & 245s & 0.62-0.67 '' & 83 & 0+8 & 3 & 0 & 5/0 & 16 \\
9 & 2024-08-17  & 245s & 1.08-1.78 '' & 70 & 4 & 4 & - & 2/- & 10 \\ 
10 - 11 & 2024-09-19  & 245s & 0.48-0.97 '' & 90 & 5+3 & 4 & 4 & 1/3 & 20 \\
12 & 2024-10-15  & 245s & 0.77-0.87 '' & 76 & 3 & - & 4 & -/3 & 10 \\ 
13 & 2024-10-17  & 245s & 0.66-0.71 '' & 88 & 0 & 2 & - & 8/- &  10 \\ 
14 - 15 & 2024-10-22  & 245s & 0.52-0.62 '' & 93 & 6+2 & 4 & 4 & 0/4 & 20 \\ 
16 - 17 & 2024-11-13  & 245s & 0.55-0.69 '' & 91 & 6+1 & 4 & 4 & 0/5 & 20 \\ 
18 & 2024-12-10  & 245s & 0.83-1.08 '' & 78 & 2 & - & 4 & -/4 & 10 \\ 
19 - 20 & 2024-12-16  & 245s & 0.67-0.90 '' & 81 & 7+2 & 3 & 4 & 0/4 & 20 \\ 
21 & 2024-12-20  & 245s & 0.60-0.73 '' & 90 & 2 & - & 4 & -/4 & 10 \\ 
22 & 2024-12-22  & 245s & 0.69-0.96 '' & 85 & 5 & 4 & - & 1/- & 10 \\ 
\hline
\multicolumn{10}{l}{*Archival observations presented in \cite{darveau-bernier_caracterisation_2023} \textdagger Mean S/N per pixel at centre of order 34 (K band)}
\end{tabular}
\end{table*}

\section{Cross--correlation template and validation}\label{app:Figs}

\begin{figure*}[h!]
    \centering
    \includegraphics[width=0.75\linewidth]{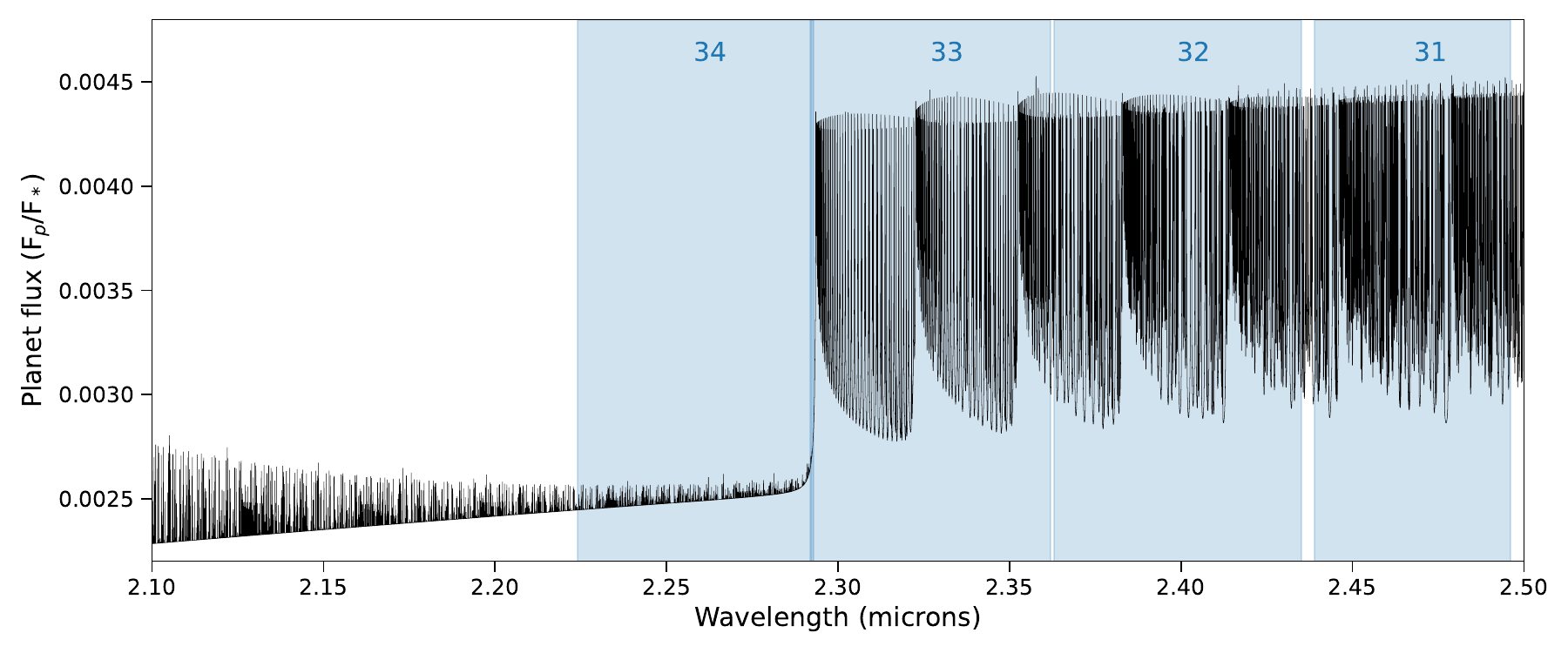}
    \caption{CO emission spectrum of WASP-33b used for the injection-recovery tests and cross--correlations presented in this work. The blue regions mark the wavelength coverage of SPIRou orders overlapping with the ro-vibrational lines of CO in the K-band.}
    \label{fig:CO_spectrum}
\end{figure*}

\begin{figure*}[h!] 
    \includegraphics[width=0.5\textwidth]{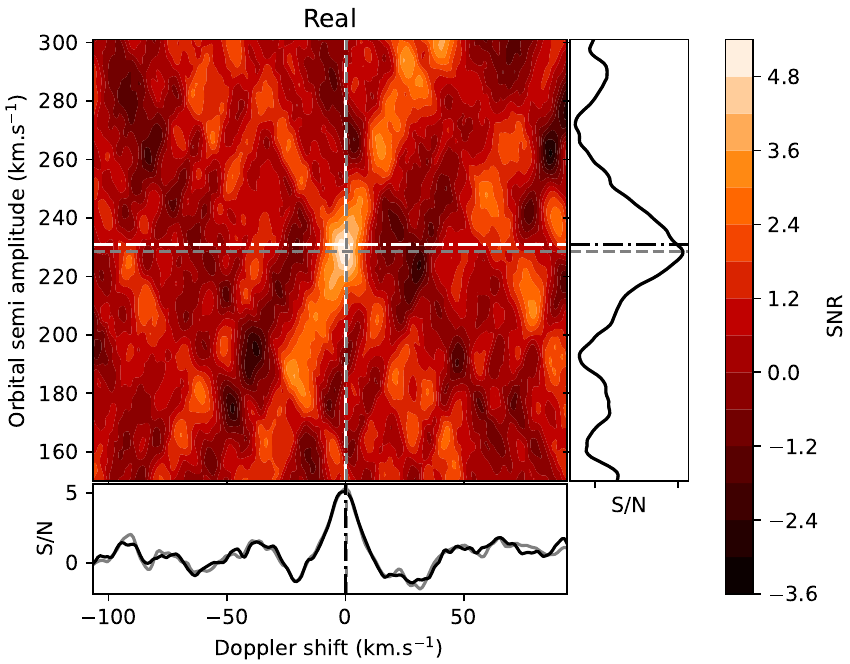}
    \includegraphics[width=0.5\textwidth]{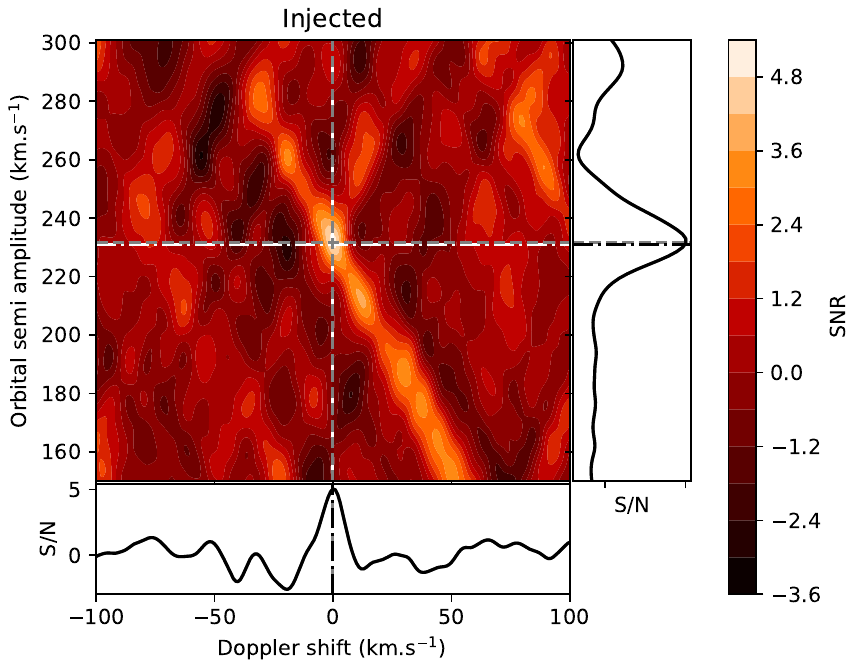}
    \caption{CO cross--correlation signal-to-noise maps from real (left) and synthetic (right) dayside observations (programme 19B), after PCA cleaning with two principal components. The expected--injected planet position is indicated by the white dash-dotted lines, while  the peak S/N from the CCF is shown as grey dashed lines. The panels on the right and bottom represent the cross--correlation S/N values along cross-sections at the expected (grey) and peak (black) K$\rm_p$ and V$\rm_{sys}$ values, respectively, with the dashed lines continuing from the white and grey lines on the full map.}
    \label{fig:KP-Vsys}
\end{figure*}

In Figure \ref{fig:CO_spectrum}, we show the (unbroadened) planetary template spectrum that we use for signal injection and recovery through CCF and/or Log-likelihood fits. Figure \ref{fig:KP-Vsys} shows how this model performs on the real dayside data of WASP-33 b and the same data where it has been injected at a velocity offset of 150 km/s. We take the similar detection strength between the two as indicative that the model is not over/underestimated as compared to the real planetary signal present in our data.

\section{Eclipse velocity offsets impact in CCF detection maps}\label{app:Model_CCF}

\begin{figure*}[ht!]
    \centering
    \includegraphics[width=\textwidth]{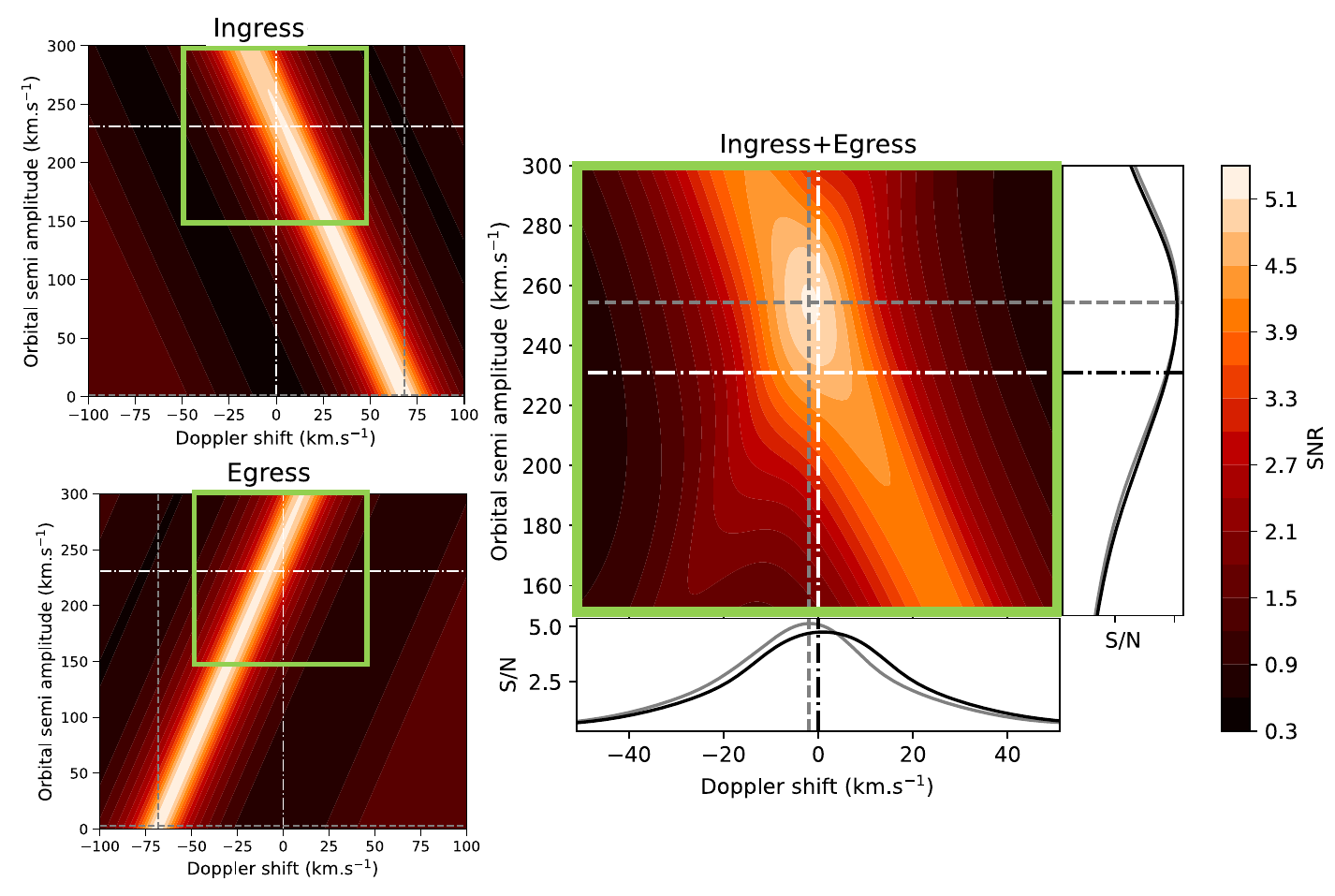}
    \caption{Model autocorrelation in K$\rm_p$-V$\rm_{sys}$ space for the injected planetary CO emission model with no noise, when considering frames from  ingress only (top left),  egress only (bottom left), and  ingress and egress (right). The combined plot is shown as a zoomed-in image compared to the individual plots, as shown by the green border. The injected model position (not including velocity offsets) is shown by the white dotted lines, while the grey dash-dotted lines indicate the location of peak S/N. The side and bottom panels in the right plot show cross-sections in K$\rm_p$ and V$\rm_{sys}$ at those same locations in black and grey, respectively.}
    \label{fig:eclipse_model_only}
\end{figure*}

Figure \ref{fig:eclipse_model_only} shows cross--correlation maps for a model with no noise, when considering only the frames inside eclipse ingress--egress and velocity offsets for a tidally locked WASP-33 b. As can be seen from the plots on the left, the CCF peaks at lower orbital velocities when fitting in ingress or egress individually, reflecting the fact that the eclipse mapping signal acts to reduce the velocity excursion (i.e. rate of change) during ingress--egress only. Since these two peaks appear at opposite systemic velocity shifts (due to the movement of the planet towards or away from us), they result in opposite systemic velocity shifts at the planet's actual orbital velocity, and thus the signals overlap at a higher orbital velocity. This effect thus results in a higher orbital velocity being preferred in the joint ingress and egress fit, with the maximum being shifted by almost 20 km/s. This highlights how velocity offsets in eclipse ingress--egress can result in very large apparent velocity offsets in detection maps, which are an artifact of the short time-coverage and summing over K$\rm_p$ and V$\rm_{sys}$ rather than a true reflection of the amplitude of these offsets.

\section{Log-likelihood results on injection-recovery tests}

\begin{figure*}[h!]
    \centering
    \includegraphics[width=0.9\textwidth]{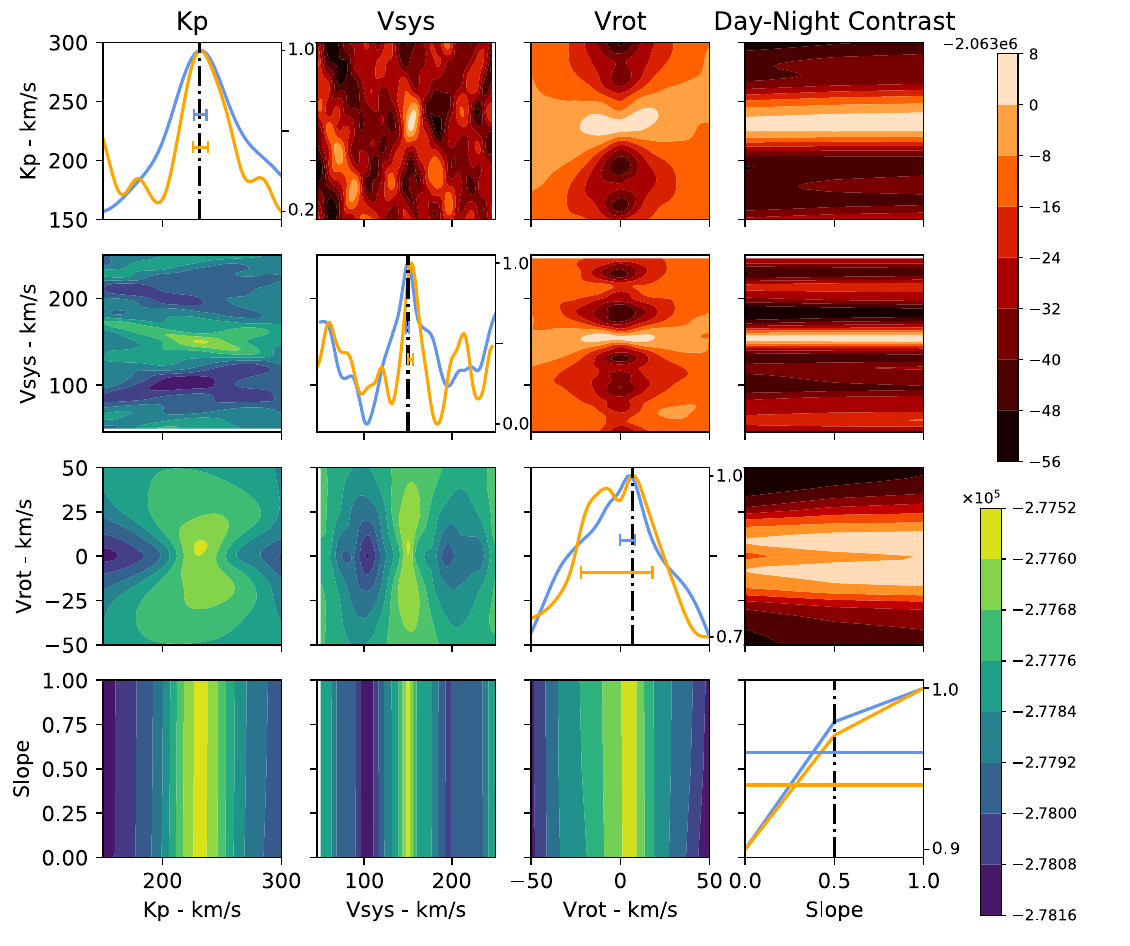}
    \caption{Full results of the Log-likelihood fit on the injected dayside data (orange) and 25 injected occultations (blue), plotted in relation to the various model parameters. Each panel is a cross-section of the final (4D) log-likelihood along one or two axes at the location of best fit. Error bars at 3$\sigma$ (computed from the $\chi^2$) are represented for each parameter in the diagonal panels. The values of the ground truth injected parameters are shown as dot-dashed lines.}\label{fig:Corner_full}
\end{figure*}

Figure \ref{fig:Corner_full} shows the results of the Log-likelihood fits on injected datasets for dayside only vs 25 ingress and egress observations.

\section{Impact of PCA in short ingress--egress time series}

\begin{figure*}[h!]
    \centering
    \includegraphics[width=\textwidth]{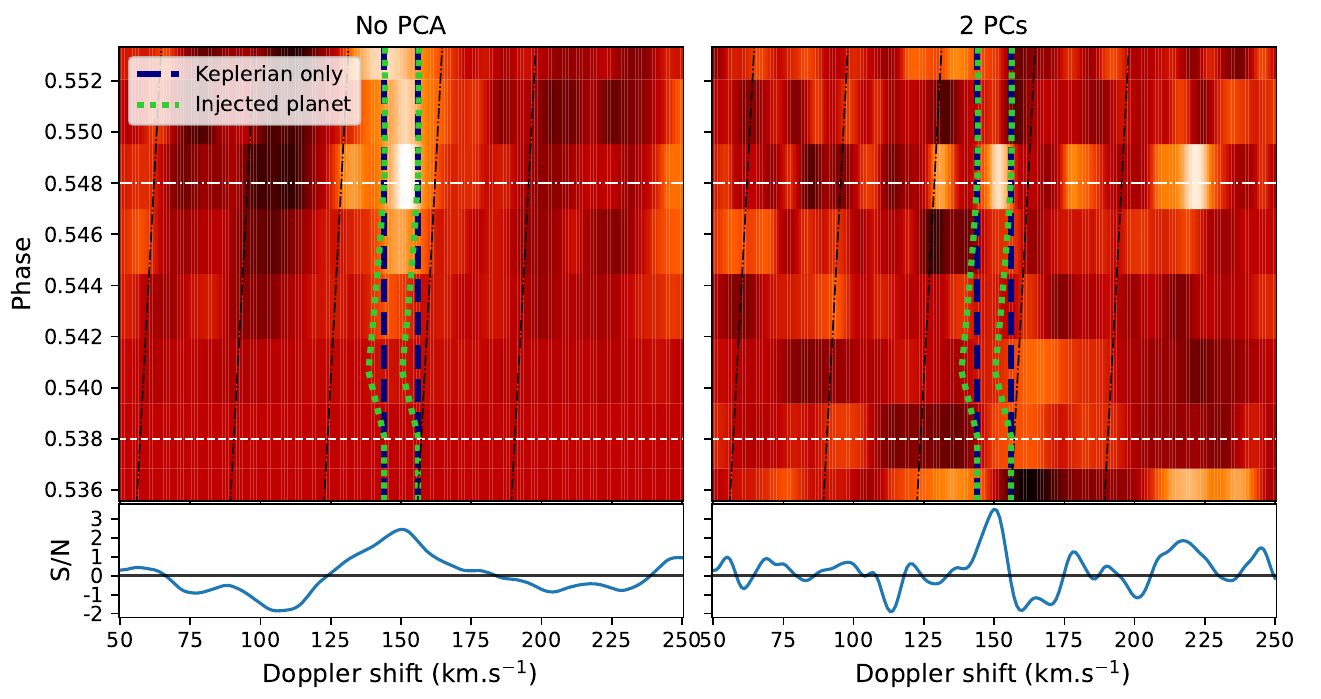}
    \caption{Planetary rest frame CCF for a synthetic egress observation with an injected planetary signal amplified by seven times ($\simeq$49 eclipses), when using no PCA (left) and two principal components (right). Between the blue dashed lines and green dotted lines are respectively the expected positions of the planetary trails if considering a simple Keplerian and the injected model with the velocity offset for a synchronous rotation of WASP-33 b. The black lines show constant velocity trails for various velocities, as could be expected from uncorrected stellar or telluric lines, and the white dashed and dash-dotted lines represent the start and end of egress, respectively. The bottom panels show the summed CCF over the entire time series. }
    \label{fig:CCF_rest_frame}
\end{figure*}

Figure \ref{fig:CCF_rest_frame} shows the unfolded CCF for a single egress in the planetary rest frame, where the signal has been amplified to bring out the planetary trail. While correlated noise at fixed velocity is visibly removed by PCA, the planetary trail is also greatly affected due to the small velocity excursion of the planet over the short time series. This shows that, while PCA does a good job at removing a lot of the amplitude in the correlated noise, it may also impacts the underlying signal by preferentially removing any slow moving components. Either way, further data is required to confirm the detection in our observed data, which may be improved through novel de-trending methods that do not suffer the same pitfalls as PCA.

\section{Doppler-eclipse mapping performance of ELT instruments}

\begin{figure*}[h!]
    \includegraphics[width=0.5\textwidth]{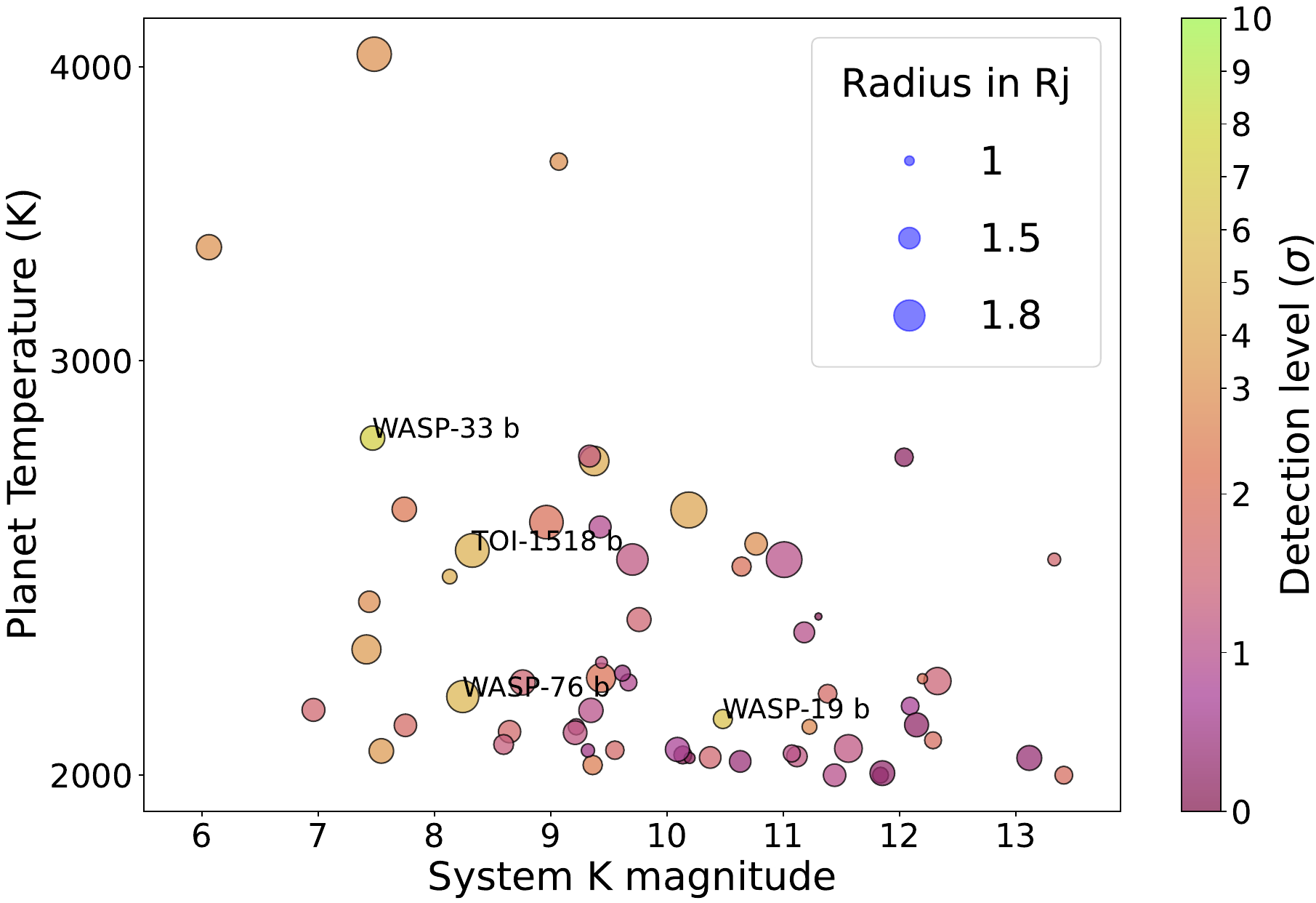}
    \includegraphics[width=0.5\textwidth]{Figures/RV_Sigma_ANDES_1occ_outline.pdf}
    \caption{Photon-limited detection S/N of the radial velocity offset induced by tidally locked rotation rates for a population of UHJs, measurable in one occultation using  ELT/MICADO (left) and ELT/ANDES (right).} \label{fig:SNR_RV_ELT}
\end{figure*}

Figure \ref{fig:SNR_RV_ELT} compares the photon-limited Doppler-Mapping performance of MICADO and ANDES on ELT, considering CO lines in the K band and a single occultation (ingress and egress). Table \ref{tab:systems} lists all the systems plotted in figures \ref{fig:SNR_RV}, \ref{fig:SNR_RV_ANDES}, \ref{fig:SNR_RV_ELT} with their computed eclipse mapping detectability, along with relevant parameters from or derived from the values obtained in the NASA exoplanet archive query (see section \ref{sec:target}).

\begin{table*}[]\label{tab:systems}
\centering
\caption{System parameters and detectability values derived and/or obtained from the NASA exoplanet archive to produce figures \ref{fig:SNR_RV}, \ref{fig:SNR_RV_ANDES}, \ref{fig:SNR_RV_ELT}.}
\resizebox{\textwidth}{!}{%
\begin{tabular}{|lcccccccc|}
\hline
\multicolumn{1}{|c|}{Planet   name} &
  \multicolumn{1}{c|}{Star Mag} &
  \multicolumn{1}{c|}{Flux Ratio} &
  \multicolumn{1}{c|}{Period} &
  \multicolumn{1}{c|}{Impact} &
  \multicolumn{1}{c|}{Synchronous Rotation} &
  \multicolumn{3}{c|}{Doppler   Eclipse Mapping Signal Detectability ($\rm\sigma$)} \\ \cline{7-9} 
\multicolumn{1}{|c|}{} &
  \multicolumn{1}{c|}{(K band)} &
  \multicolumn{1}{c|}{F$_p$/F$_*$   (2.3 $\rm\mu$m)} &
  \multicolumn{1}{c|}{(days)} &
  \multicolumn{1}{c|}{Parameter} &
  \multicolumn{1}{c|}{Velocity (km/s)} &
  \multicolumn{1}{l|}{ANDES - 1 eclipse} &
  \multicolumn{1}{l|}{MICADO   - 1 eclipse} &
  \multicolumn{1}{l|}{SPIROU   - 10 eclipses} \\ \hline
WASP-189 b        & 6.1  & 1.05E-02 & 2.72 & 0.46  & 3.04  & 7.0  & 2.2 & 1.9 \\ \hline
HD 202772 A b     & 7.0  & 3.51E-03 & 3.31 & 0.41  & 2.39  & 3.9  & 1.1 & 1.0 \\ \hline
KELT-20 b         & 7.4  & 8.90E-03 & 3.47 & 0.50  & 2.56  & 8.6  & 2.5 & 2.3 \\ \hline
TOI-1431 b        & 7.4  & 5.97E-03 & 2.65 & 0.89  & 2.87  & 6.6  & 2.0 & 1.8 \\ \hline
WASP-33 b         & 7.5  & 1.96E-02 & 1.22 & 0.22  & 6.67  & 10.5 & 5.2 & 3.1 \\ \hline
KELT-9 b          & 7.5  & 1.52E-02 & 1.48 & 0.18  & 6.53  & 4.4  & 2.2 & 1.3 \\ \hline
KELT-7 b          & 7.5  & 6.03E-03 & 2.73 & 0.59  & 2.99  & 7.9  & 2.5 & 2.2 \\ \hline
MASCARA-1 b       & 7.7  & 7.79E-03 & 2.15 & 0.11  & 3.80  & 4.5  & 1.6 & 1.3 \\ \hline
MASCARA-4 b       & 7.8  & 4.30E-03 & 2.82 & 0.14  & 2.77  & 4.2  & 1.3 & 1.1 \\ \hline
WASP-18 b         & 8.1  & 1.23E-02 & 0.94 & 0.38  & 6.73  & 6.8  & 3.4 & 2.0 \\ \hline
WASP-76 b         & 8.2  & 1.16E-02 & 1.81 & 0.14  & 5.17  & 9.1  & 3.8 & 2.7 \\ \hline
TOI-1518 b        & 8.3  & 1.14E-02 & 1.90 & 0.84  & 5.04  & 8.6  & 3.6 & 2.5 \\ \hline
HD 2685 b         & 8.6  & 6.67E-03 & 4.13 & 0.10  & 1.78  & 3.4  & 0.9 & 0.9 \\ \hline
KELT-17 b         & 8.6  & 6.06E-03 & 3.08 & 0.57  & 2.53  & 4.0  & 1.2 & 1.1 \\ \hline
WASP-82 b         & 8.8  & 6.06E-03 & 2.71 & 0.16  & 3.06  & 3.4  & 1.1 & 0.9 \\ \hline
HAT-P-70 b        & 9.0  & 1.08E-02 & 2.74 & -0.62 & 3.48  & 4.3  & 1.4 & 1.2 \\ \hline
TOI-2109 b        & 9.1  & 2.27E-02 & 0.67 & 0.71  & 10.24 & 3.2  & 2.1 & 1.0 \\ \hline
KELT-18 b         & 9.2  & 5.65E-03 & 2.87 & 0.10  & 2.79  & 2.7  & 0.8 & 0.7 \\ \hline
TOI-1181 b        & 9.2  & 5.57E-03 & 2.10 & 0.22  & 3.16  & 2.3  & 0.7 & 0.6 \\ \hline
WASP-71 b         & 9.3  & 3.88E-03 & 2.90 & 0.43  & 2.08  & 1.3  & 0.4 & 0.3 \\ \hline
HAT-P-7 b         & 9.3  & 1.10E-02 & 2.20 & 0.49  & 3.50  & 2.6  & 0.9 & 0.7 \\ \hline
HAT-P-49 b        & 9.3  & 5.15E-03 & 2.69 & 0.34  & 3.02  & 2.4  & 0.7 & 0.6 \\ \hline
WASP-3 b          & 9.4  & 7.83E-03 & 1.85 & 0.44  & 3.93  & 4.6  & 1.6 & 1.3 \\ \hline
WASP-121 b        & 9.4  & 2.50E-02 & 1.27 & 0.10  & 7.03  & 6.5  & 3.4 & 1.9 \\ \hline
KOI-13 b          & 9.4  & 9.24E-03 & 1.76 & 0.24  & 4.38  & 1.7  & 0.6 & 0.5 \\ \hline
HAT-P-57 b        & 9.4  & 9.47E-03 & 2.47 & 0.18  & 3.61  & 4.4  & 1.5 & 1.2 \\ \hline
KELT-1 b          & 9.4  & 6.19E-03 & 1.22 & 0.14  & 4.66  & 2.0  & 0.8 & 0.6 \\ \hline
WASP-87 b         & 9.6  & 6.02E-03 & 1.68 & 0.60  & 4.21  & 3.3  & 1.2 & 0.9 \\ \hline
WASP-72 b         & 9.6  & 3.88E-03 & 2.22 & 0.26  & 2.97  & 1.0  & 0.3 & 0.3 \\ \hline
WASP-100 b        & 9.7  & 6.63E-03 & 2.85 & 0.63  & 2.39  & 2.2  & 0.6 & 0.6 \\ \hline
WASP-178 b        & 9.7  & 9.90E-03 & 3.34 & 0.54  & 2.77  & 2.8  & 0.8 & 0.7 \\ \hline
WASP-167 b        & 9.8  & 8.45E-03 & 2.02 & 0.77  & 3.99  & 3.0  & 1.1 & 0.8 \\ \hline
KELT-21 b         & 10.1 & 6.10E-03 & 3.61 & 0.42  & 2.24  & 1.9  & 0.5 & 0.5 \\ \hline
Kepler-91 b       & 10.1 & 6.98E-04 & 6.25 & 0.86  & 1.12  & 0.2  & 0.0 & 0.0 \\ \hline
WASP-12 b         & 10.2 & 2.30E-02 & 1.09 & 0.34  & 9.07  & 4.8  & 3.0 & 1.5 \\ \hline
Kepler-1658 b     & 10.2 & 1.18E-03 & 3.85 & 0.94  & 1.42  & 0.3  & 0.1 & 0.1 \\ \hline
WASP-48 b         & 10.4 & 8.42E-03 & 2.14 & 0.73  & 3.58  & 3.3  & 1.1 & 0.9 \\ \hline
WASP-19 b         & 10.5 & 2.25E-02 & 0.79 & 0.66  & 9.17  & 7.2  & 4.4 & 2.2 \\ \hline
TOI-4329 b        & 10.6 & 3.81E-03 & 2.92 & 0.00  & 2.62  & 0.9  & 0.3 & 0.3 \\ \hline
KELT-16 b         & 10.6 & 1.63E-02 & 0.97 & 0.31  & 7.46  & 2.6  & 1.4 & 0.8 \\ \hline
WASP-103 b        & 10.8 & 1.86E-02 & 0.93 & 0.19  & 8.44  & 3.5  & 2.1 & 1.1 \\ \hline
WASP-78 b         & 11.0 & 9.76E-03 & 2.18 & 0.42  & 4.53  & 1.9  & 0.7 & 0.6 \\ \hline
WASP-114 b        & 11.1 & 8.25E-03 & 1.55 & 0.45  & 4.42  & 2.0  & 0.7 & 0.6 \\ \hline
HATS-35 b         & 11.1 & 8.89E-03 & 1.82 & 0.26  & 4.11  & 2.3  & 0.8 & 0.7 \\ \hline
GPX-1 b           & 11.2 & 9.29E-03 & 1.74 & 0.81  & 4.31  & 1.9  & 0.7 & 0.5 \\ \hline
TOI-1937 A b      & 11.2 & 1.41E-02 & 0.95 & 0.86  & 6.73  & 3.8  & 1.9 & 1.1 \\ \hline
Kepler-1517 b     & 11.3 & 2.18E-03 & 5.55 & 0.83  & 0.80  & 0.1  & 0.0 & 0.0 \\ \hline
HATS-24 b         & 11.4 & 1.67E-02 & 1.35 & 0.33  & 5.29  & 2.9  & 1.3 & 0.9 \\ \hline
WASP-142 b        & 11.4 & 7.50E-03 & 2.05 & 0.77  & 3.81  & 2.0  & 0.7 & 0.6 \\ \hline
Qatar-7 b         & 11.6 & 1.01E-02 & 2.03 & 0.08  & 4.28  & 2.2  & 0.8 & 0.6 \\ \hline
CoRoT-19 b        & 11.8 & 5.16E-03 & 3.90 & 0.23  & 1.69  & 0.7  & 0.2 & 0.2 \\ \hline
K2-52 b           & 11.9 & 3.56E-03 & 3.54 & 0.33  & 2.32  & 0.7  & 0.2 & 0.2 \\ \hline
HATS-70 b         & 12.0 & 7.52E-03 & 1.89 & 0.24  & 3.75  & 0.4  & 0.1 & 0.1 \\ \hline
Kepler-76 b       & 12.1 & 1.02E-02 & 1.54 & 0.00  & 4.50  & 1.3  & 0.5 & 0.4 \\ \hline
HATS-40 b         & 12.1 & 4.40E-03 & 3.26 & 0.35  & 2.47  & 0.6  & 0.2 & 0.2 \\ \hline
Wendelstein-1   b & 12.2 & 6.07E-02 & 2.66 & 0.77  & 1.98  & 5.2  & 1.4 & 1.4 \\ \hline
HATS-18 b         & 12.3 & 1.83E-02 & 0.84 & 0.30  & 8.11  & 2.3  & 1.3 & 0.7 \\ \hline
HATS-67 b         & 12.3 & 1.36E-02 & 1.61 & 0.86  & 5.35  & 2.4  & 1.0 & 0.7 \\ \hline
OGLE2-TR-L9 b     & 13.1 & 8.06E-03 & 2.49 & 0.00  & 3.31  & 0.7  & 0.2 & 0.2 \\ \hline
Wendelstein-2   b & 13.3 & 7.95E-02 & 1.75 & 0.31  & 3.38  & 3.4  & 1.1 & 0.9 \\ \hline
WTS-2 b           & 13.4 & 3.88E-02 & 1.02 & 0.59  & 6.84  & 2.6  & 1.3 & 0.8 \\ \hline
OGLE-TR-132 b     & 15.0 & 6.90E-03 & 1.69 & 0.56  & 3.63  & 0.4  & 0.1 & 0.1 \\ \hline
OGLE-TR-56 b      & 16.6 & 7.18E-03 & 1.21 & 0.62  & 5.75  & 3.4  & 1.5 & 1.0 \\ \hline
\end{tabular}%
}
\end{table*}

\end{appendix}

\end{document}